\documentclass[a4paper,fleqn,usenatbib,useAMS]{mnras}

\usepackage{graphicx}	% Including figure files
\usepackage{amsmath}	% Advanced maths commands
\usepackage{multicol}        % multicolumn entries in tables
\usepackage{pdflscape}	% Landscape pages

\usepackage{ latexsym,graphicx,natbib,eufrak,url, hyperref}

\DeclareMathAlphabet{\mathcal}{OMS}{cmsy}{m}{n}

\def\limepy{{\textsc{limepy}}}
\def\scipy{{\textsc{scipy}}}
\def\numpy{\textsc{numpy}}
\def\matplotlib{\textsc{matplotlib}}
\def\python{\textsc{python}}
\def\integrate{\textsc{integrate}}

\newcommand{\rvec}{\textbf{\emph{r}}}
\newcommand{\vvec}{\textbf{\emph{v}}}

\newcommand{\fone}{\tfrac{1}{2}}
\newcommand{\fthree}{\tfrac{3}{2}}
\newcommand{\ffive}{\tfrac{5}{2}}
\newcommand{\fseven}{\tfrac{7}{2}}

\def\dr{{\rm d}}

\def\hyp{{_1}F_1}

\def\ehat{\hat{E}}

\def\Eg{E_\gamma}
\def\Etot{E_{\rm tot}}

\def\gdos{\mathfrak{g}}

\def\Ncomp{N_{\rm comp}}

\def\threehalf{\textstyle{\frac{3}{2}}\displaystyle}
\def\fivehalf{\textstyle{\frac{5}{2}}\displaystyle}

\def\gthree{g+\threehalf}
\def\gfive{g+\fivehalf}
\def\henon{H\'{e}non}
\def\kms{{\rm km}\,{\rm s}^{-1}}
\def\rmm{r_{\rm m}}
\def\rh{r_{\rm h}}
\def\rhhat{\hat{r}_{\rm h}}
\def\rhat{\hat{r}}
\def\rrahat{\hat{p}}
\def\rahat{\hat{r}_{\rm a}}
\def\rajhat{\hat{r}_{{\rm a},j}}
\def\rv{r_{\rm v}}
\def\rvhat{\hat{r}_{\rm v}}

\def\ra{r_{\rm a}}
\def\rs{r_{\rm s}}
\def\ms{M_{\rm s}}
\def\rt{r_{\rm t}}

\def\rthat{\hat{r}_{\rm t}}
\def\pc{{\rm pc}}
\def\phit{\phit_{\rm t}}
\def\phat{\hat{p}}
\def\phihat{\hat{\phi}}
\def\rhohat{\hat{\rho}}

\def\rhoint{\mathcal{I}^{\rho}}

\def\fintone{\mathcal{I}_{F,1}}
\def\finttwo{\mathcal{I}_{F,2}}
\def\rhointnull{\mathcal{I}_0^{\rho}}
\def\vsqint{\mathcal{I}^{\rho \sigma^2}}
\def\vrsqint{\mathcal{I}^{\rho \sigma_{\rm r}^2}}
\def\vtsqint{\mathcal{I}^{\rho \sigma_{\rm t}^2}}
\def\mvsq{\sigma^2}
\def\sigmaoned{\sigma_{\rm 1d}}
\def\sigmaonedj{\sigma_{{\rm 1d}, j0}}
\def\sigmahat{\hat{\sigma}}
\def\sigmathat{\hat{\sigma}_{\rm t}}
\def\sigmarhat{\hat{\sigma}_{\rm r}}
\def\sigmat{\sigma_{\rm t}}
\def\sigmar{\sigma_{\rm r}}
\def\sigmatheta{\sigma_\theta}
\def\sigmaphi{\sigma_\varphi}

\def\rvec{\textbf{\emph{r}}}
\def\vvec{\textbf{\emph{v}}}
\def\mhat{\hat{M}}
\def\Uhat{\hat{U}}

\def\khat{\hat{k}}
\def\vr{v_{\rm r}}
\def\vt{v_{\rm t}}
\def\vtheta{v_\theta}
\def\vphi{v_\varphi}

\def\msun{\mbox{M}_{\rm \odot}}

\usepackage[T1]{fontenc}
\usepackage{ae,aecompl}

\usepackage{txfonts}
\DeclareMathAlphabet{\mathpzc}{OT1}{pzc}{m}{it}

\title[A family of lowered isothermal models]
  {A family of lowered isothermal models}
\author[Mark Gieles and Alice Zocchi]
  {Mark Gieles\thanks{E-mail: \href{mailto:m.gieles@surrey.ac.uk}{m.gieles@surrey.ac.uk} (MG)} and Alice Zocchi\thanks{E-mail: \href{mailto:a.zocchi@surrey.ac.uk}{a.zocchi@surrey.ac.uk} (AZ)}\\
Department of Physics, University of Surrey, Guildford, GU2 7XH, UK.\\
}
\date{Accepted 2015 August 7.  Received 2015 August 6; in original form 2015 June 26}
\def\LaTeX{L\kern-.36em\raise.3ex\hbox{a}\kern-.15em
    T\kern-.1667em\lower.7ex\hbox{E}\kern-.125emX}

\begin{document}         
\maketitle
\begin{abstract}
We present a family of self-consistent, spherical, lowered isothermal
models, consisting of one or more mass components, with parameterized
prescriptions for the energy truncation and for the amount of radially
biased pressure anisotropy. The models are particularly suited to
describe the phase-space density of stars in tidally limited,
mass-segregated star clusters in all stages of their life-cycle. The
models extend a family of isotropic, single-mass models by
Gomez-Leyton and Velazquez, of which the well-known Woolley, King and
Wilson (in the non-rotating and isotropic limit) models are
members. We derive analytic expressions for the density and velocity
dispersion components in terms of potential and radius, and introduce
a fast model solver in \python\ (\limepy), that can be used for data
fitting or for generating discrete samples.
\end{abstract}

%%%%%%%%%%%%%%%%%%%%%%%%%%%%%%%%%%%%%%%%%%%%
\section{Introduction}
\label{Sect:Intro}
The evolution of globular clusters (GCs) is the result of an interplay
between stellar astrophysics (stellar and binary evolution, stellar
mergers, etc.), dynamical two-body relaxation and the interaction with
the tidal field of their host
galaxy \citep{2003gmbp.book.....H}. Despite this plethora of physical
processes at work on their respective time-scales, the {\it
instantaneous} surface brightness profiles and kinematics of GCs are
well described by relatively simple distribution function (DF) based
models that need very few
assumptions \citep*{1979AJ.....84..752G,1997A&ARv...8....1M,
2012A&A...539A..65Z}.

The relative simplicity of GC properties is owing to the absence of
gas and non-baryonic dark matter and the collisional nature of their
evolution, which drives them to tractable properties, such as
spherical symmetry, isotropy and (quasi-)equipartition between
different mass species \citep[e.g.][]{1987degc.book.....S}.  Because
the relaxation time-scale of GCs is much longer than their dynamical
time, their instantaneous properties can be described by models that
satisfy the collisionless Boltzmann equation \citep[see e.g. Chapter 8
in][]{2014dyga.book.....B}.

Two-body interactions in GCs evolve the velocity distribution of stars
towards a Maxwell-Boltzmann distribution, at least in the core, where
the relaxation time-scale is short. Models with isothermal cores are
therefore a good choice for fitting properties of GCs. An obvious
starting point for a discussion on model choice is, therefore, the
isothermal model. This model has an infinite spatial extent and
infinite mass \citep{1939isss.book.....C} and to make the model
applicable to real star clusters, the assumption of the idealized
Maxwell-Boltzmann distribution of velocities needs to be relaxed. This
can be done by changing the model such that stars have a finite escape
velocity. \citet{1954MNRAS.114..191W} developed such a model by simply
`lowering' the (specific) energy $E$ by a constant. The DF, which
describes the density in six-dimensional phase-space as a function of
$E$, is then simply $f(E) = A\exp[-(E-\phi(\rt))/s^2]$, for
$E\le\phi(\rt)$, and $f(E)=0$ for $E>\phi(\rt)$. Here $s$ is a
velocity scale, which in the isothermal model equals the one-dimensional
velocity dispersion and $E$ is reduced by the specific potential at
the truncation radius $\rt$, $\phi(\rt)$.  This truncation in energy
mimics the role of tides due to the host galaxy, which makes it easier
for stars to escape by reducing the escape velocity. The resulting
models are nearly isothermal in the core, and have a finite mass and
extent.

The  DF of these models is discontinuous at
 $E=\phi(\rt)$. \citet{1963MNRAS.125..127M}
 and \citet{1966AJ.....71...64K} avoided this  by subtracting
 a constant from the DF introduced by Woolley, which makes the models
 continuous at $E=\phi(\rt)$.  Compared to the Woolley models, the
 density of stars near the escape energy is reduced in these models
 (hereafter referred to as King models), and they display a more
 gentle truncation of their density profile. The resulting, more
 extended, low-density envelopes make these models resemble real GCs
 more \citep[for an in depth discussion on the effect of the
 truncation on the density profiles see][]{1977AJ.....82..271H}.  The
 spherical, non-rotating limit of the models introduced
 by \citet{1975AJ.....80..175W}, hereafter called Wilson models, are
 models that are continuous both in the DF and its derivative. This is
 achieved by subtracting an additional term linear in $E$ from the
 DF. These models are yet more spatially extended than King
 models. For some GCs in Local Group galaxies, the Wilson models
 provide a better description of the observed surface brightness
 profiles compared to the King models
 (\citealt{2005ApJS..161..304M}; \citealt{2012MNRAS.419...14C} also show that models that are more extended than King models better describe the surface brightness profiles of some GCs).
 
An additional outcome of the two-body relaxation process is that it
drives the velocity distribution of the stars towards
isotropy. Isotropic models, defined by a DF that only depends on $E$,
are therefore a natural choice for clusters that are in late stages of
their evolution, near dissolution. At early phases, however, the
velocity distribution in the outer parts is expected to be radially
anisotropic. This is, first, because the (incomplete) violent
relaxation process that takes place during their formation results in
a halo of radial orbits \citep{1967MNRAS.136..101L}. Secondly,
two-body ejections from the dense core populate the halo with radial
orbits on a two-body relaxation
time-scale \citep{1972ApJ...173..529S}.  \citet{1963MNRAS.125..127M}
proposed a separable DF, dependent on $E$ and on the (specific)
angular momentum $J$ to introduce radial anisotropy (hereafter
referred to as Michie-King models). The DF of the Michie-King models
is the product of the isotropic DF with an exponential term with a
$J^2$ dependent argument. This is similar to Eddington's method of
including radial anisotropy in the isothermal
model \citep{1915MNRAS..75..366E}. As a result, the inner parts of the
models remain approximately isothermal and isotropic, which is
appropriate to GCs because there the relaxation time is short, and
anisotropy becomes important at larger distances from the centre. Near
the truncation radius the models become isotropic again as a result of
the energy truncation. The latter property has a somewhat coincidental
resemblance to GCs, because near the Jacobi radius the orbits of stars
gain angular momentum due to the interaction with the (tri-axial)
 tidal potential \citep{1992ApJ...386..519O}, therewith
suppressing the amount of radial anisotropy near the truncation
energy. A review of the effect of anisotropy on model properties can
be found in \citet{1982ARA&A..20..399B}.

In real GCs, which contain multiple mass components, the relaxation
process drives the systems towards equipartition, resulting in the
heavier components being more centrally concentrated, a state which is
often referred to as mass segregated.  King models with different mass
species were first introduced by \citet{1976ApJ...206..128D} and have
since been applied to take into account the effects of mass
segregation in mass-modelling efforts of Galactic GCs
(e.g. M3: \citealt{1979AJ.....84..752G}, Omega
Cen: \citealt{1987A&A...184..144M} and larger samples of
GCs: \citealt{1993ASPC...50..357P, 2012ApJ...755..156S}). Mass
segregation is important for almost all of the Galactic GCs, given
their short relaxation time-scales, relative to their ages \citep*{H61,
2011MNRAS.413.2509G}. Approximating multimass systems by single-mass
models can lead to severe biases in the inferred properties of
GCs \citep{2015MNRAS.448L..94S, sollima15} and it is, therefore,
desirable to have the ability to include multiple mass components in a
dynamical model of a GC.

It is our aim to develop a family of models that capture the general
behaviour of collisional systems discussed above, and whose properties
can be varied by parameters that can be constrained by observational
data. \citet{1977A&A....61..391D} showed that the expressions for the
DF of the isotropic Woolley, King and Wilson models can be generalized
by a DF in which the exponential function of $E$ is reduced by the
leading orders of its series expansion. This approach was further
generalized by \citet[][hereafter GV14]{2014JSMTE..04..006G}, who
showed that solutions {\it in between} these models can be obtained
(these models are briefly reviewed in Section~\ref{ssec:iso}). In this
paper we extend the models of GV14 to allow for the presence of
(radially biased) pressure anisotropy and multiple mass components. We
present an efficient Poisson solver in \python\ to facilitate the use
of these models in fitting observational data, and in drawing samples
from the models, which can be used as initial conditions for numerical
simulations.

The paper is organized as follows: in Section~\ref{sec:model}, we
define the models and in Section~\ref{sec:properties}, we illustrate
their main properties.  In Section~\ref{sec:limepycode}, we present the
code \limepy\footnote{\limepy\ is available
from \url{https://github.com/mgieles/limepy.}} and our conclusions and
a discussion are presented in Section~\ref{sec:conclusion}. Supporting
material can be found in the appendices.

%%%%%%%%%%%%%%%%%%%%%%%%%%%%%%%%%%%%%%%%%%%%
\section{Model definition and scaling}
\label{sec:model}
\subsection{Single-mass models}
\subsubsection{Distribution function (DF)}
\label{ssec:df}
The DF of  the single-mass family of models is
\begin{equation}
 f(E,J^2) =
\displaystyle A\exp\left(-\frac{J^2}{2\ra^2 s^2}\right)\Eg\left(g, -\frac{E-\phi(\rt)}{s^2}\right)
\label{eq:dfani}
\end{equation}
for $E\le\phi(\rt)$, and 0 for $E>\phi(\rt)$.  The DF depends on two
integrals of motion: the specific energy $E = v^2/2 + \phi(r)$, with
$v$ the velocity and $\phi(r)$ the specific potential at distance $r$
from the centre, and the norm of the specific angular momentum vector
$J = |\rvec \times \vvec|=rv\sin\vartheta$, where $\rvec$ and $\vvec$
are the position vector and velocity vector, respectively, and
$\vartheta$ is the angle between them. The energy $E$ is lowered by
the potential at the truncation radius $\phi(\rt)$.

In equation~(\ref{eq:dfani}) we introduced the function
\begin{align}
\Eg(a, x) =  
\begin{cases}
\exp(x)  &  a=0 \\
\displaystyle\exp(x) P(a, x)  &  a>0,
\end{cases}
\label{eq:eg}
\end{align}
where $P(a, x) \equiv \gamma(a, x)/\Gamma(a)$ is the regularized lower
incomplete gamma function (see Appendix~\ref{app:gamma} for the
definition of this function and its properties). Combining the
exponential and the incomplete gamma function into a single function
$\Eg(a, x)$ has advantages in deriving the model properties (see GV14
and Appendix~\ref{AppD:Eg} for details on the behaviour of this
function).

A model is specified by three parameters: the central potential, which
is a required boundary condition for solving Poisson's equation and
defines how concentrated the model is; the anisotropy radius $\ra$,
which determines the amount of anisotropy present in the system (for
increasing $\ra$ the models are more isotropic); the truncation
parameter $g$, which controls the sharpness of the truncation of the
model (this parameter is called $\gamma$ in GV14). The physical units
of a model are defined by two scales: the velocity scale $s$, and the
normalization constant $A$, which sets the phase-space density and
therewith the total mass $M$. For more information regarding scales and parameters of
the models we refer the reader to Section~\ref{ssec:scaling}.

The isotropic models ($\ra\rightarrow\infty$) and their properties are
discussed in detail in GV14. For these models, and integer values of
$g$, three well-known families of models are recovered: when $g=0$ we
retrieve the \citet{1954MNRAS.114..191W} models, for $g=1$ we recover
the King models \citep[][]{1963MNRAS.125..127M,1966AJ.....71...64K},
and for $g=2$ we find the (isotropic, non-rotating) Wilson
models \citep[][]{1975AJ.....80..175W}\footnote{The Woolley, King and
Wilson DFs follow straightforwardly from equations~(\ref{eq:dfani})
and (\ref{eq:eg}), because $\Eg(0,x) = \exp(x)$, $P(1,x) =
1-\exp(-x)$, such that $\Eg(1,x) = \exp(x)-1$ and $P(2,x) =
1-\exp(-x)-x\exp(-x)$, such that $\Eg(2,x) = \exp(x) - 1 - x$.}. In
practice, the models defined by equation~(\ref{eq:dfani}) are radially
anisotropic for $\ra\lesssim\rt$, because of the $J^2$ dependence in
the first exponential. When $g=1$, the DF is the Michie-King
model \citep{1963MNRAS.125..127M}, which is often used to fit GC
data \citep[e.g.][]{1987A&A...184..144M, 2012ApJ...755..156S}.

The potential $\phi(r)$ is found by solving Poisson's equation. For
the self-consistent problem we consider here, the potential is
completely determined by the density $\rho$ associated with the
DF. This problem is non-linear, because the DF depends on the
potential. Since the models defined by equation~(\ref{eq:dfani}) are
spherically symmetric, Poisson's equation is
\begin{align}
\frac{1}{r^2}\frac{\dr}{\dr r}\left(r^2\frac{\dr\phi}{\dr r}\right)& = 4\pi G\rho,
\end{align}
where the density is obtained by means of an integration of the DF over all velocities 
\begin{equation}
\rho = \int  \dr^3 v \, f(E, J^2).
\label{eq:rho}
\end{equation}
In Sections~\ref{ssec:iso} and \ref{ssec:ani}, we derive analytic
expressions for $\rho$ as a function of $\phi$ and $r$.  Note that
only in the anisotropic case the dependence on the radial coordinate
$r$ is both implicit (through $\phi$, as in the isotropic case), and
explicit, i.e. $\rho(\phi, r)$.  Having analytic expressions for
$\rho(\phi, r)$, avoids the need of solving a double integral at each
radial step, making it significantly faster to obtain the solution to
Poisson's equation.  In the next section we introduce a convenient set
of units to solve the model.

%_________________________________________________________
\subsubsection{Scaling and units}
\label{ssec:scaling}
To solve Poisson's equation, we use a dimensionless (positive) energy
$\ehat = \phihat - \khat$, with dimensionless potential $\phihat =
(\phi(\rt) - \phi)/s^2$, and $\khat \equiv v^2/(2 s^2)$. As
in \citet{1966AJ.....71...64K}, we consider the dimensionless density
by normalizing $\rho$ to its central value, i.e. $\rhohat
= \rho/\rho_0$. In this way, Poisson's equation in dimensionless form
reads
\begin{equation}
\frac{1}{\rhat^2}\frac{\dr}{\dr\rhat}\left(\rhat^2\frac{\dr\phihat}{\dr\rhat}\right) = -9\rhohat.
\end{equation}
The dimensionless radius is now defined by the other scales: $\rhat =
r/\rs$, with $\rs^2 = 9 s^2/(4\pi G\rho_0)$. This radial scale was
introduced in \citet{1966AJ.....71...64K} and is often referred to as
the King radius. The factor of 9 was introduced to give $\rs$ the
meaning of a core radius, because for models with moderately high
central concentration, the projected density at $\rs$ is about one
half of its central value.

The Poisson equation is solved by assuming the boundary conditions at
$\rhat = 0$: $\phihat = \phihat_0$ and $\dr\phihat/\dr r = 0$. As
mentioned in Section~\ref{ssec:df}, the central potential $\phihat_0$
is one of the parameters that define the model\footnote{This parameter
is called $W_0$ in \citet{1966AJ.....71...64K}.}.

\subsubsection{Isotropic models}
\label{ssec:iso}
We first briefly review the isotropic version of these models, as
introduced by GV14. Many quantities can be derived from the DF. The
density $\rho$ is found by integrating the DF over all velocities
(equation~\ref{eq:rho}) and the pressure is found by taking the second
velocity moment of the DF
\footnote{By considering the first velocity moment of the DF we find the mean velocity: for these models, this quantity vanishes everywhere. We also note that expressions for higher order moments of the velocity distribution can be derived, but these are beyond the scope of this paper. }
\begin{align}
\rho &= (2\pi s^2)^{3/2}A\rhoint, \label{eq:rhodef} \\
\rho\mvsq &= (2\pi s^2)^{3/2}s^2A \vsqint.
\end{align}
Here $\sigma^2=3\sigmaoned^2$ is the mean-square velocity,
$\sigmaoned$ is the one-dimensional velocity dispersion and we introduce
a dimensionless density integral ($\rhoint$) and a dimensionless
pressure integral ($\vsqint$)
\begin{align}
\rhoint &= \frac{2}{\sqrt{\pi}}\int_0^{\phihat} \dr \khat \, \khat^{1/2}\Eg(g, \phihat - \khat) = \Eg(\gthree, \phihat), \label{eq:rhointiso}\\
\vsqint &= \frac{4}{\sqrt{\pi}}\int_0^{\phihat} \dr \khat \, \khat^{3/2}\Eg(g, \phihat - \khat) = 3\Eg(\gfive, \phihat). \label{eq:vsqintiso}
\end{align}
The results of these integrations follow straightforwardly from the
convolution formula of the $\Eg(a,x)$ function
(equation~\ref{EG_convolution}). An alternative derivation by means of
fractional calculus is presented in Appendix~\ref{app:fractional}.
The dimensionless density that appears in Poisson's equation is
therefore $\rhohat=\rhoint/\rhointnull$, where $\rhointnull$ is the
result of equation~(\ref{eq:rhointiso}) evaluated at
$\phihat=\phihat_0$. The dimensionless mean-square velocity is found
from $\sigmahat^2 = \sigma^2/s^2 = \vsqint/\rhoint$.

%_________________________________________________________
\subsubsection{Anisotropic models}
\label{ssec:ani}
Here we present the relevant quantities for the anisotropic case.  The
details of the derivations can be found in Appendix~\ref{app:series},
and the derivations by means of fractional calculus can be found in
Appendix~\ref{app:fractional}.  To solve the anisotropic models, we
introduce $t=\cos\theta$, such that we can write the integral over the
angles as $4\pi\int_0^1 \dr t$. We further introduce $\rrahat= \rhat/\rahat$ such that the density integral becomes

\begin{align}
\rhoint&=\frac{2}{\sqrt{\pi}}\!\int_0^{\phihat}\!\dr\khat\int_{0}^{1}\! \dr t \, \exp\left[\khat \phat^2 (t^2-1)\right] \khat^{1/2}\Eg(g, \phihat - \khat)  \nonumber \\
 &=\frac{2}{\sqrt{\pi}} \int_0^{\phihat}\dr\khat\, \frac{F(\phat\khat^{1/2})}{\phat}\Eg(g, \phihat - \khat).
\label{eq:rhointani}
\end{align}
Here $F(x)$ is Dawson's integral and we refer to
Appendix~\ref{App:Dawson} for some properties of this function. To
first order, $F(x) \propto x$, and we thus find that for large
$\rahat$, i.e. small $\phat$, equation~(\ref{eq:rhointani}) converges
to the integral of the isotropic model
(equation~\ref{eq:rhointiso}). The solution of the integration gives
 $\rhoint$ as a function of $\phihat$ and $\rrahat$
\begin{equation}
\rhoint \!=\! \frac{\Eg( \gthree, \phihat)}{1+\phat^2}+\frac{\phat^2}{1+\phat^2}   \frac{\phihat^{g+\fthree}{_1}F_1(1, \gfive, -\phihat\phat^2)}{\Gamma(g+\ffive)}.
\label{eq:rhointanires}
\end{equation}
Here $\hyp(a,b, x)$ is the confluent hypergeometric function whose
properties are given in Appendix~\ref{1F1}. For small $\phat$, the
second term on the right-hand-side goes to zero and the solution
converges to the isotropic result of
equation~(\ref{eq:rhointiso}). This  expression for the
density integral allows for fast computations of the right-hand-side
of Poisson's equation and facilitates efficient solving of the
anisotropic models.

For the anisotropic models, we need to calculate both the radial and
the tangential\footnote{The tangential velocity comprises the two
components $\vt^2 = \vtheta^2+\vphi^2$, where $\vtheta =
v\sin\theta\sin\varphi$ and $\vphi = v\sin\theta\cos\varphi$. The
corresponding components of the velocity dispersion tensor are equal
to each other, and each of them accounts for half of the tangential
component: $\sigmat^2 = 2 \sigmatheta^2 = 2\sigmaphi^2$.}  components
of the pressure tensor, as well as the total pressure.  The radial and
tangential component of the velocity vector are defined as $\vr =
v\cos\theta$ and $\vt = v\sin\theta$ and for the corresponding
integrals we find
\begin{align}
\displaystyle\vrsqint&\hspace{-0.cm}=\hspace{-0.cm}\frac{4}{\sqrt{\pi}}\!\int_0^{\phihat}\!\!\!\dr\khat\!\int_{0}^{1}\!\!\!\dr t\,\exp\!\left[\rrahat^2\khat(t^2\!\!-\!1)\right]t^2\khat^{3/2}\Eg(g, \phihat-\khat),\\
\displaystyle\vtsqint&\hspace{-0.cm}=\hspace{-0.cm}\frac{4}{\sqrt{\pi}}\!\int_0^{\phihat}\!\!\!\dr\khat\!\int_{0}^{1}\!\!\!\dr t\,\exp\!\left[\rrahat^2\khat(t^2\!-\!1)\right](1\!-t^2)\khat^{3/2}\Eg(g, \phihat-\khat),\\
\displaystyle\vsqint&\hspace{-0.cm}=\hspace{-0.cm}\frac{4}{\sqrt{\pi}}\!\int_0^{\phihat}\!\!\!\dr\khat\!\int_{0}^{1}\!\!\!\dr t\,\exp\!\left[\rrahat^2\khat(t^2\!-\!1)\right]\khat^{3/2}\Eg(g, \phihat-\khat).
\end{align}
By carrying out these integrals as described in Appendices~\ref{sapp:sigma_deriv} and \ref{sapp:sigma_deriv_frac}, we  obtain
\begin{align}
\hspace{-0.25cm}\vrsqint &\!=\!\frac{\Eg(\gfive, \phihat)}{1+\phat^2}\!+\!\frac{\phat^2}{1+\phat^2} \frac{\phihat^{g+\ffive}\hyp(1, g\!+\!\fseven, \!-\phihat\phat^2)}{\Gamma(g+\fseven)}, \label{eq:vrsqintres} \\
\hspace{-0.25cm}\vtsqint &= \frac{\Eg(\gfive, \phihat)}{1+\phat^2}\frac{2}{(1+\phat^2)} + \frac{2\phat^2}{1+\phat^2}\frac{\phihat^{g+\ffive}}{\Gamma(g+\fseven)} \nonumber\\
                 &\times\left[  \frac{\hyp(1, g+\fseven, -\phihat\phat^2)}{1+\phat^2} + \hyp(2, g+\fseven, -\phihat\phat^2) 	\right],  \label{eq:vtsqintres} \\
\hspace{-0.25cm}\vsqint&\!=\!\frac{\Eg(\gfive, \phihat)}{1+\phat^2}\frac{(3+\phat^2)}{(1+\phat^2)} +\frac{\phat^2}{1+\phat^2} \frac{\phihat^{g+\ffive}}{\Gamma(g+\fseven)} \nonumber\\
                \times&\left[\!\frac{3+\phat^2}{1+\phat^2}{_1}F_1(1, g\!+\!\fseven,\!-\phihat\phat^2)\!+\!2{_1}F_1(2, g\!+\!\fseven,\!-\phihat\phat^2)\right]\!. \label{eq:vsqintres} 
\end{align}
Note that the expression for $\vrsqint$ resembles the expression for
$\rhoint$ of equation~(\ref{eq:rhointanires}), in the sense that the
functional form is the same, but all arguments and the power index
that include $g$ are increased by 1. We already saw a similar
resemblance between $\rhoint$ and $\vsqint$ in the isotropic case
(equations~\ref{eq:rhointiso} and \ref{eq:vsqintiso}, respectively).

With these expressions for the density and pressure integrals, we
defined most of the properties of these models that are of direct
relevance for comparison to data. In Section~\ref{ssec:project}, we
discuss how the projected quantities can be derived.

%_________________________________________________________
\subsubsection{Limits}
\label{sssec:limits}
In this section we consider some limits of the models.  In the core,
where $\phat$ is small ($\rhat\ll\rahat$), the model is isotropic. This
is because the second terms in equations~(\ref{eq:rhointanires}),
(\ref{eq:vrsqintres}), (\ref{eq:vtsqintres}) and (\ref{eq:vsqintres})
vanish due to the multiplication by $\phat^2$.

Near the truncation radius the models  behave like polytropes and are, therefore, also isotropic, because 
\begin{align}
\lim_{\phihat\rightarrow0} \hyp(1, a\!+\!1,-\phat^2\phihat) &= 1,\\
\lim_{\phihat\rightarrow0} \Eg(a, \phihat) &= \frac{\phihat^a}{\Gamma(a+1)},
\end{align}
and the $\rrahat$ dependence disappears. In this regime, we  find
\begin{align}
\lim_{\phihat\rightarrow0} \rhohat & = \frac{\phihat^{g+3/2}}{\Gamma(g+5/2)}, \label{eq:poly}\\
\lim_{\phihat\rightarrow0} \rhohat \sigmahat^2 & = 3\frac{\phihat^{g+5/2}}{\Gamma(g+7/2)},\\
\lim_{\phihat\rightarrow0} \rhohat \sigmarhat^2 & = \frac{1}{3}\lim_{\phihat\rightarrow0} \rhohat \sigmahat^2 = \frac{1}{2} \lim_{\phihat\rightarrow0} \rhohat \sigmathat^2.
\end{align}
This suppression of the velocity anisotropy near the truncation radius
results naturally from the mathematical definition of the truncation,
and is appropriate for tidally truncated
systems \citep{1992ApJ...386..519O}. In $N$-body models a tangentially
biased anisotropy is observed near $\rt$ \citep{sollima15}, which
cannot be reproduced by the models presented here. However, it is
likely that most of the stars with tangentially biased velocities are
above the escape energy, so-called potential escapers and these are
not considered by these models, nor any other model we are aware off.

Models with $\phihat_0\rightarrow0$ are close to pure polytropes over
their entire radial range. In this regime, and for $g=7/2$ (i.e. a
polytropic index $n=5$, equation~\ref{eq:poly}), we recover
the \citet{1911MNRAS..71..460P} model, which is infinite in extent
($\rho \propto r^{-5}$ at large radii), but finite in mass. Polytropes
with $n\ge 5$ (i.e.  $g\ge7/2$) are infinite in extent and will not be
considered here. For $g<7/2$ models can have a finite $\rt$ depending
on both $\phihat_0$ and $\ra$ (see GV14 and
Section~\ref{sec:properties}).

In the cores of models with $\phihat_0\gg0$ the DF approaches the
isothermal sphere, because
\begin{align}
\lim_{\phihat\rightarrow\infty} \Eg(a,\phihat) & =\exp(\phihat).
\end{align}
Models with $g\rightarrow\infty$ also approach the isothermal
sphere. To conclude, these models approach the isothermal sphere in
the limit of $\phihat_0\rightarrow\infty$, independent of $g$, but
also in the limit of $g\rightarrow\infty$, independent of $\phihat_0$.

%_________________________________________________________
\subsection{Multimass models}
\label{ssec:multimassmodel}
It is possible to consider models with multiple mass components, by
considering the DF as the sum of DFs of the form of
equation~(\ref{eq:dfani}), each of which describes a different mass
component with a mass-dependent velocity scale parameter. The first to
do this were \citet{1976ApJ...206..128D}, who calculated multimass
King models.  For a multimass model with $\Ncomp$ mass components,
$2\Ncomp+2$ parameters are required in addition to the ones introduced
in Section~\ref{ssec:df} for single-mass models. These additional
parameters are the values for the component masses $m_j$, the amount
of mass in each component $M_j$, $\delta$ and $\eta$. The latter two
parameters set the mass dependence of the velocity scale $s_j$ and
the anisotropy radius of each component $\rajhat$, for which we adopt
power-law relations
\begin{align}
s_j &= s\mu_j^{-\delta}, \label{eq:delta} \\
\rajhat &= \rahat\mu_j^{\eta}. \label{eq:eta}
\end{align}
Here $\mu_j= m_j/\bar{m}$ is the dimensionless mass of component $j$
and $\bar{m}$ is the central density weighted mean-mass
\begin{equation}
\bar{m} = \frac{\sum_j m_j\rho_{0j}}{\sum_j \rho_{0j}}.
\end{equation}
Note that in the multimass models, the values of $s$ and $\rahat$ are
the velocity scale and anisotropy radius corresponding to $\bar{m}$.
The definitions of $\delta$ and $\eta$ are such that the anisotropy
profiles are approximately mass independent when $\delta=\eta$ (see
equation~\ref{eq:dfani}). The typical values considered for these
parameters are $\delta = 1/2$ and $\eta = 0$.

We notice that in the limit of infinite $\phihat_0$ the velocity scale
$s_j$ approaches the one-dimensional velocity dispersion of mass
component $j$, $\sigma_{{\rm 1d},j}$, hence the traditional assumption for $\delta=1/2$
implies equipartition ($m_j s_j^2 =\bar{m}\sigma_{{\rm 1d},j}^2=$
constant). However, it is important to keep in mind that for
multimass models with typical and realistic values of $\phihat_0$,
the velocity dispersion of each component in the centre is smaller
than $s_j$ and, therefore, there is no equipartition (see
Section~\ref{ssec:delta} and \citealt{1981AJ.....86..318M,
2006MNRAS.366..227M}).

To solve a multimass model self-consistently, we compute the density
for each mass component as in equation~(\ref{eq:rho}) and add all
components on the right-hand-side of Poisson's equation. The detailed
procedure is described in \citet{1979AJ.....84..752G}, and here we
only briefly summarize the required steps. The dimensionless Poisson
equation to solve is
\begin{equation}
\hat{\nabla}^2\phihat = -9\sum_j\alpha_j\rhohat_j,
\end{equation}
where $\alpha_j$ is the ratio of the central density of the $j$-th
mass component to the total central density, such that
\begin{equation}
\sum_j \alpha_j = \sum_j \frac{\rho_{0j}}{\rho_0} =  1 
\end{equation}
and 
\begin{equation}
\rhohat_j = \frac{\rho_j}{\rho_{0j}} = \frac{\rhoint(\mu^{2\delta}\phihat,\hat{r})}{{\rhoint(\mu^{2\delta}\phihat_0, 0)}}.
\label{eq:rhointratio}
\end{equation}
By considering multiple mass components, we introduce an eigenvalue
problem in the solution of Poisson's equation, because the values of
$\rho_{0j}$ that yield the desired ${M}_j$ values are not known a
priori. Therefore, as a first step to solve the model, we assume that
$\alpha_j = {M}_j/\sum_j M_j$, and we obtain the solution by iteration
(see Section~\ref{sec:limepycode} for details).

%___________________________________________
\subsection{Normalization and potential energy}
In solving the models we have chosen to define the dimensionless
quantities in terms of the density scale $\rho_0$ and the velocity
scale $s$ (Section~\ref{ssec:scaling}). In some cases it is useful to
have an expression for the normalization constant $A$ in the DF
(equation~\ref{eq:dfani}), for example, when fitting models to
discrete data. From equation~(\ref{eq:rhodef}) we find that $A$
relates to the other scales as

\begin{align}
A    & = \frac{\rho_0}{(2\pi s^2)^{3/2}\rhointnull}.
\end{align}
For the multimass models there is a normalisation for each component,
$A_j$. The relation with the mass scale $\ms=M/\mhat$ is
$\ms=\rs^3\rho_0=\rs^3(2\pi s^2)^{3/2} A\rhointnull$, where we
introduced $\mhat = \int\rhohat\dr^3\rhat$.

The total dimensionless (positive) gravitational energy $\Uhat$ of the
model is calculated from integrating the
potential \citep{1966AJ.....71...64K}
\begin{align}
\Uhat &= \frac{1}{2} \int_{0}^{\mhat}  \dr \hat{m} \, \phihat +  \frac{\hat{G}\mhat^2}{2\rthat}.
\label{eq:Uhat}
\end{align}
The second term has to be added because $\phihat$ is a lowered
potential.  Note that this integration of $\phihat$ over mass is
readily obtained from solving Poisson's equation.

%%%%%%%%%%%%%%%%%%%%%%%%%%%%%%%%%%%%%%%%%%%%
\section{Model properties}
\label{sec:properties}

\subsection{Single-mass models}
\subsubsection{Density and velocity dispersion profiles}
\begin{figure}
\includegraphics[width=\columnwidth]{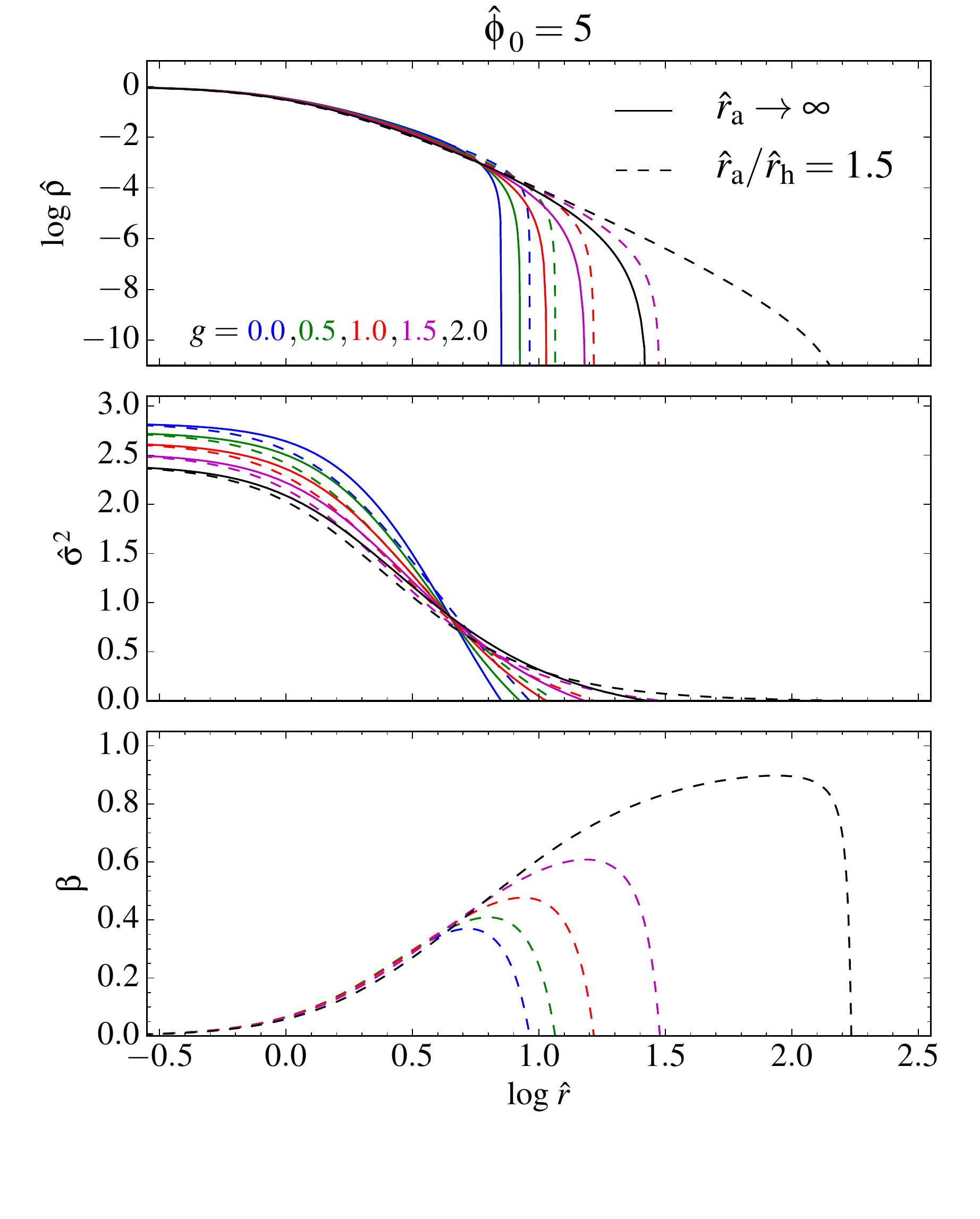}
 \caption{Dimensionless density profile (top), velocity dispersion
 profile (middle) and anisotropy profile (bottom) for models with
 different truncation parameters $g$ (different colours). Isotropic
 models are shown with solid lines, anisotropic models with
 $\rahat/\rhhat=1.5$ with dashed lines.}
\label{fig:rhovel}
\end{figure}

In Fig.~\ref{fig:rhovel}, we show the density profiles, the velocity
dispersion profiles, and the anisotropy profiles for isotropic and
anisotropic models with different values of the truncation parameter
$g$. The anisotropy profile is computed from $\sigmat^2$ and
$\sigmar^2$ as
\begin{equation}
\beta = 1-\frac{\sigmat^2}{2 \sigmar^2}.
\label{eq:beta}
\end{equation}
In the case of isotropy $\beta = 0$, $0<\beta\le1$ indicates radially
biased anisotropy (with $\beta=1$ implying fully radial orbits) and
for tangentially biased anisotropy $\beta<0$. Because $\beta$ is a
measure of anisotropy locally, we also quantify the total amount of
anisotropy with
\begin{equation}
\kappa = \frac{2K_{\rm r}}{K_{\rm t}},
\label{eq:kappa}
\end{equation}
introduced by \citet{1981SvA....25..533P}. Here $K_{\rm r}$ and
$K_{\rm t}$ are the radial and tangential components of the kinetic
energy, respectively.  For isotropic models $\kappa =1$, and for
radially biased anisotropic models
$\kappa>1$. \citet{1981SvA....25..533P} found that for
$\kappa>1.7\pm0.25$ radial orbit instability occurs.  We use this
criterion to check the stability of the anisotropic models we
calculate.

In Fig.~\ref{fig:rhovel} we show anisotropic models characterized by
$\rahat/\rhhat=1.5$. Because the (dimensionless) half-mass radius
$\rhhat$ is not known before solving the model, we find the value of
$\rahat$ that gives the correct ratio $\rahat/\rhhat$ iteratively. We
see that all models are approximately isothermal in the centre. When
increasing $g$, the models become more extended. Including radial
anisotropy also results in a larger truncation radius.

Note that, with this choice of $\rahat/\rhhat$, the maximum value
assumed by the anisotropy function for $g = 0$ (Woolley model) is
about $0.4$, while for $g = 2$ (Wilson model) it is possible to
achieve $\beta\simeq1$ in the outer parts of the model. This
dependence of the maximum value of $\beta$ on $g$ does not imply that
there are differences in the total amount of anisotropy: for all the
anisotropic models shown in Fig.~\ref{fig:rhovel}, indeed, we find
$\kappa\simeq1.2$. The ability to calculate models with more radial
orbits (larger $\beta$) without increasing the radial component of the
total kinetic energy is important to keep in mind when considering
other physical effects that can enhance or suppress the amount of
radial orbits, such as the presence of a dark matter
halo \citep{2013MNRAS.428.3648I} and the galactic
tides \citep{1992ApJ...386..519O}. In a forthcoming study, we quantify
the presence of radial orbits in direct $N$-body models of tidally
limited clusters \citep{2016MNRAS.462..696Z}.

%______________________________________________________________________
\subsubsection{DF, density of states and differential energy distribution}
\begin{figure*}
\includegraphics[width=\textwidth]{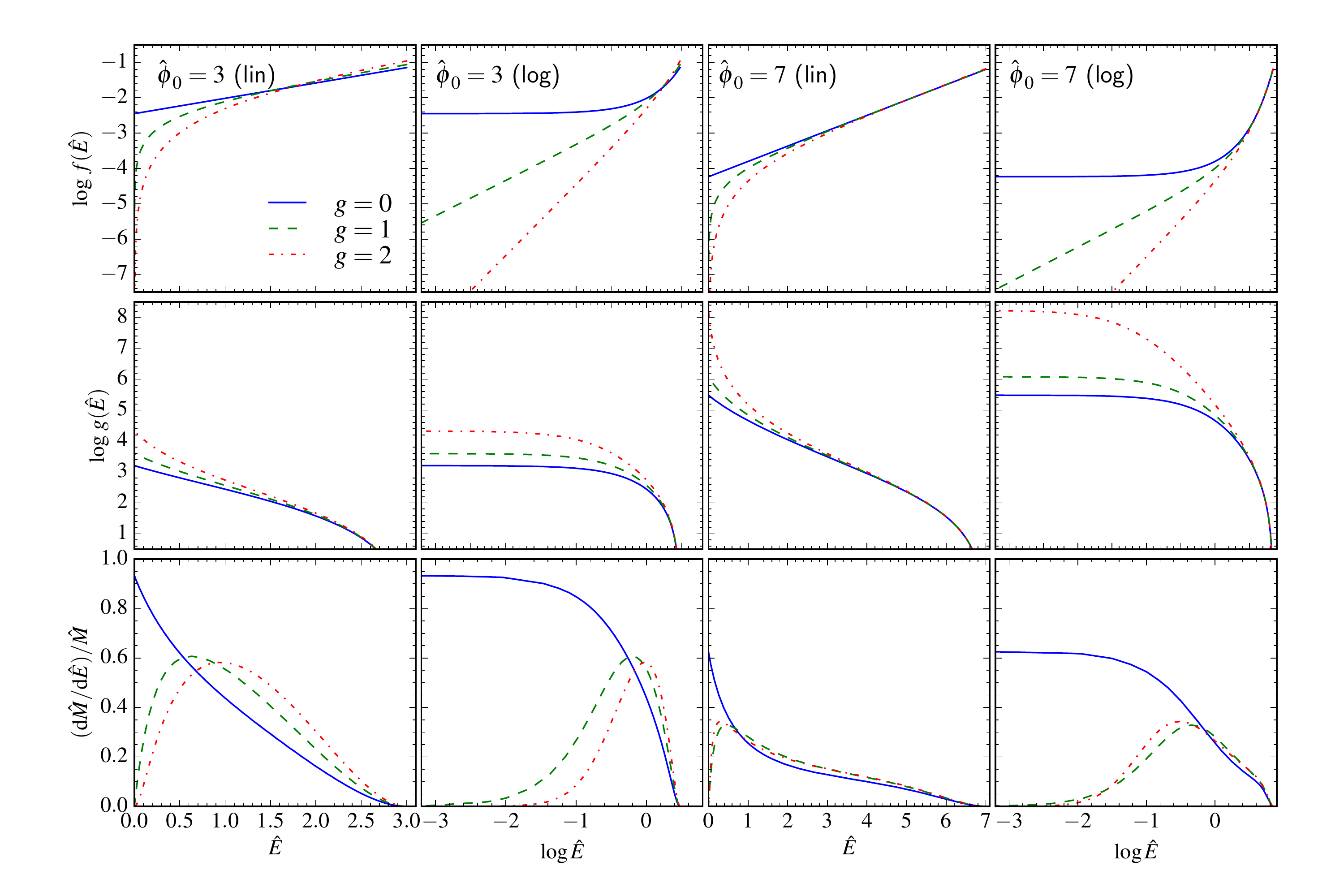}
 \caption{Distribution function (DF, $f(\ehat)$) (top), the density of
 states $\gdos(\ehat)$ (middle) and the normalized differential
 energy distribution $(\dr \mhat/\dr\ehat)/\hat{M} =
 f(\ehat)\gdos(\ehat)/\mhat$ (bottom) for isotropic models with
 different central potentials ($\phihat_0 = 3$ in the two left columns
 and $\phihat_0 = 7$ in the two right columns) and different $g$ (see
 the legend in the top left panel). }
\label{fig:dmde_comp}
\end{figure*}

In the top panels of Fig.~\ref{fig:dmde_comp}, we show the DF as a
function of $\ehat$, for isotropic models, with different values of
$g$ and $\phihat_0$. In the middle panels we show the density of
states $\gdos(\ehat)$, which is the phase-space volume per unit of
energy (see equation~\ref{densityofstates} for a definition). The
bottom panels display the differential energy distribution
$\dr \mhat/\dr \ehat$, which is the amount of mass per unit
energy. For the isotropic models it is simply the product of
$f(\ehat)$ and $\gdos(\ehat)$ (equation~\ref{eq:dmde}). Details on how
this was derived for the models presented here, and on the procedure
for anisotropic models, are given in Appendix~\ref{app:dmde}. A
general discussion on the differential energy distribution can be
found in chapter~4 of \citet{BT1987}.

In the first and third columns (linear $x$-scale), we recognize the
exponential behaviour of $f(\ehat)$ for the $g=0$ model, and the
exponential behaviour at high $\ehat$ for $g>0$ models.  From the
second and fourth column, we see that at low $\ehat$, the DF scales as
$f(\ehat) \propto \ehat^g$, which corresponds to the regime where the
models behave as polytropes.  From Fig.~\ref{fig:dmde_comp} it is also
evident that when $\ehat \simeq \phihat_0$, the model behaviour is
independent of $g$.

From the differential energy distribution, we see that only for $g=0$
there is a non-zero mass at $\ehat=0$. For models with $g>0$, the
truncation is such that
$f(\ehat=0)=\dr \hat{M}/\dr \ehat\left|_{\ehat=0}\right.=0$. These
models give rise to more realistic looking density profiles, but in
real GCs the number of particles with the escape energy is not
zero \citep{2001MNRAS.325.1323B}, because of the gradual scattering of
particles over the critical energy for escape by two-body relaxation,
and because of the finite time for stars to escape from the Jacobi
surface imposed by the galactic tidal
field \citep{2000MNRAS.318..753F}.

%______________________________________________________________________
\subsubsection{Finite and infinite models}

As discussed in Section~\ref{sssec:limits}, there are no models with
finite extent if $g\ge3.5$. GV14 showed that the maximum value $g_{\rm
max}$ to get models with a finite extent depends on $\phihat_0$, and
$g_{\rm max}=3.5$ holds in the limit of $\phihat_0\rightarrow0$. GV14
show that all their isotropic models are finite for $g \lesssim 2.1$.

\begin{figure}
\includegraphics[width=\columnwidth]{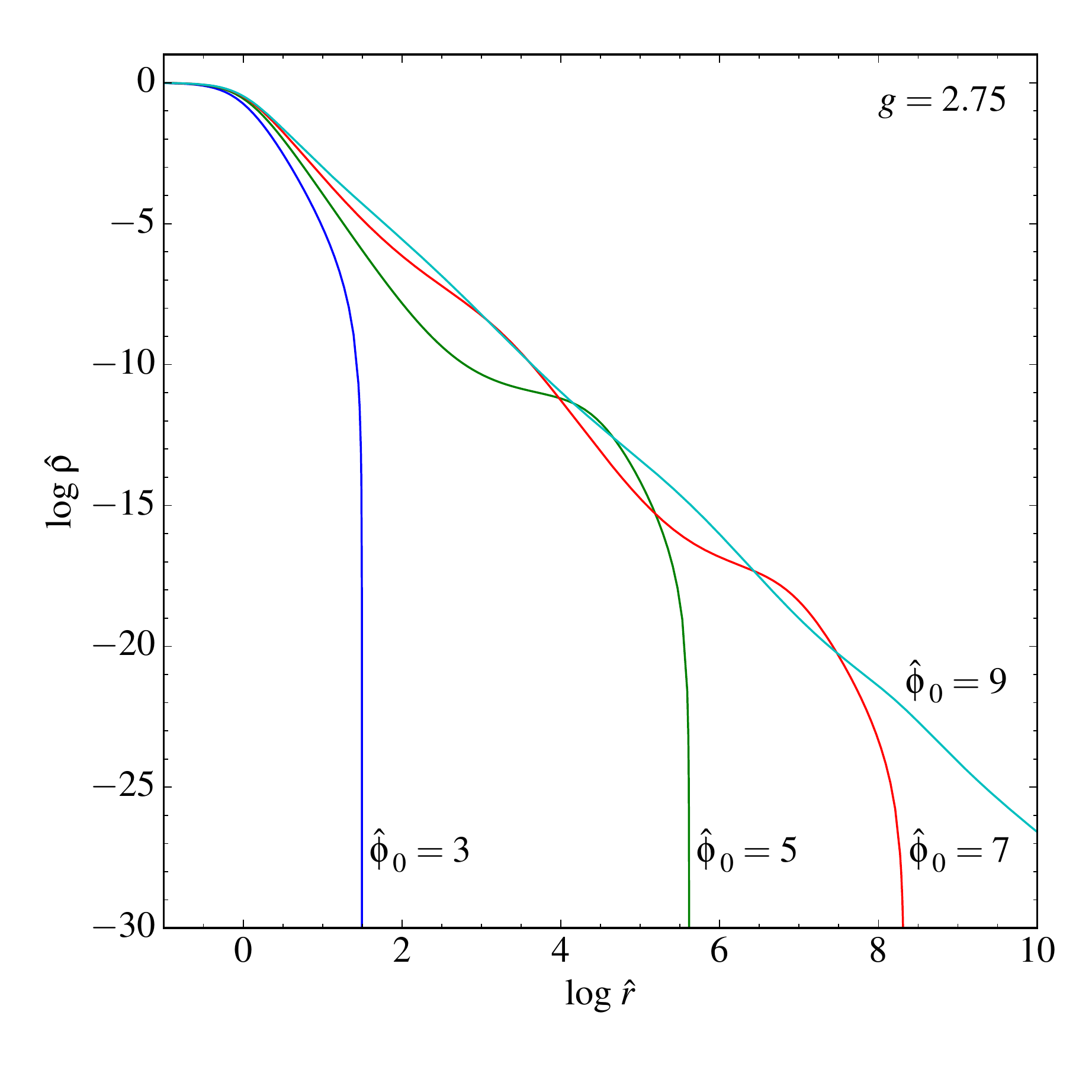}
 \caption{Density profiles for isotropic models with truncation
 parameter $g=2.75$. Models with $\phihat_0 = 3$, $5$, and $7$ (blue,
 green, and red line, respectively) have a finite truncation radius,
 but only the model with $\phihat_0=3$ is relevant when describing
 GCs; the model with $\phihat_0=9$ (light blue line) is infinite in
 extent.}
\label{fig:rho_multi}
\end{figure}

\begin{figure}
\includegraphics[width=\columnwidth]{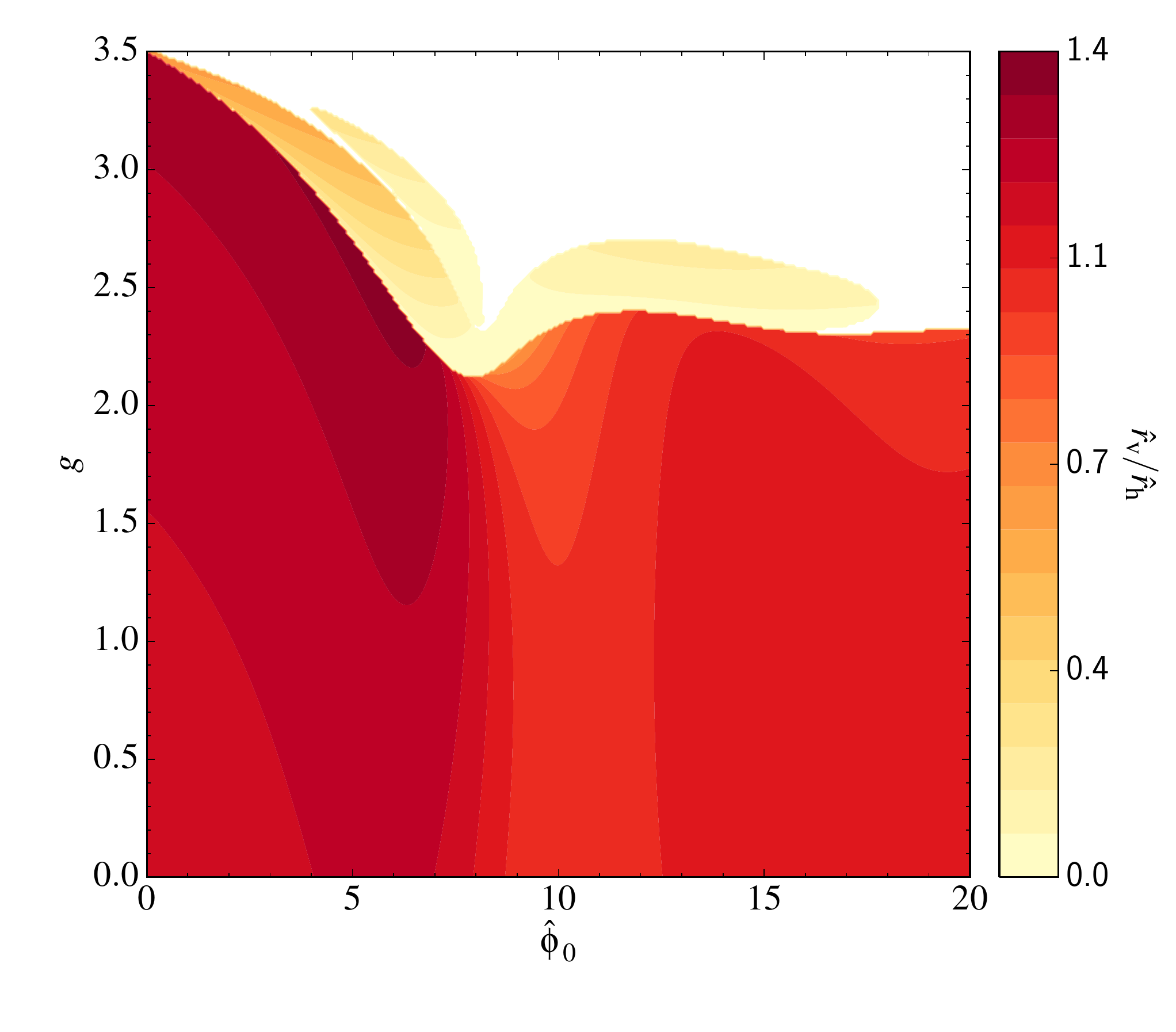}
 \caption{Ratio of dimensionless virial radius to half-mass radius, $\rvhat/\rhhat$,
 for models with different $\phihat_0$ and $g$. We consider models
 with $\rvhat/\rhhat\ge0.64$ as relevant to describe star
 clusters. Models that have an infinite $\rthat$ are plotted as
 $\rvhat/\rhhat=0$ (i.e. they correspond to the white region in the
 plot).}
\label{fig:rvrh_phi0}
\end{figure}

We note that there is a class of isotropic models that are finite in
extent, but are not relevant to star clusters, and that are not
discussed in GV14. This is illustrated in Fig.~\ref{fig:rho_multi},
where we show density profiles for models with different $\phihat_0$
and $g=2.75$. The model with $\phihat_0=3$ converges to a finite
$\rthat$ and has a density profile comparable in shape to the ones
shown in Fig.~\ref{fig:rhovel}. The model with $\phihat_0=9$ is
infinite in extent, and only plotted up to $\log\,\rhat=10$. The
models with $\phihat_0=5$ and $\phihat_0=7$ are finite, but show a
sharp upturn in the density profile at large radii, which causes them
to have a lot of mass in the envelope, but little energy, which makes
these models inapplicable to real stellar systems.  Their extreme
density contrast between the core and the extended halo makes these
models perhaps applicable to red giant stars \citep[see the density
profiles for red giants in ][]{2012ApJ...744...52P}. To quantify the
boundary between models with, and without the core-halo structure, we
compute the ratio of the dimensionless virial radius $\rvhat=-G\mhat^2/(2\Uhat)$
over $\rhhat$ for a grid of models with $0\leq\phihat_0\leq20$ and
$0\leq g\leq3.5$, and we show the result as contours in
Fig.~\ref{fig:rvrh_phi0}. We find that for a given $g(\phihat_0)$,
when increasing $\phihat_0(g)$, the change in $\rvhat/\rhhat$ is large
and abrupt once the models develop the core-halo structure. We
identify the value of $\rvhat/\rhhat\simeq0.64$ as the one separating
the two classes of models. In the remaining discussion, we only
consider models with $\rvhat>0.64\rhhat$.

\begin{figure}
\includegraphics[width=8cm]{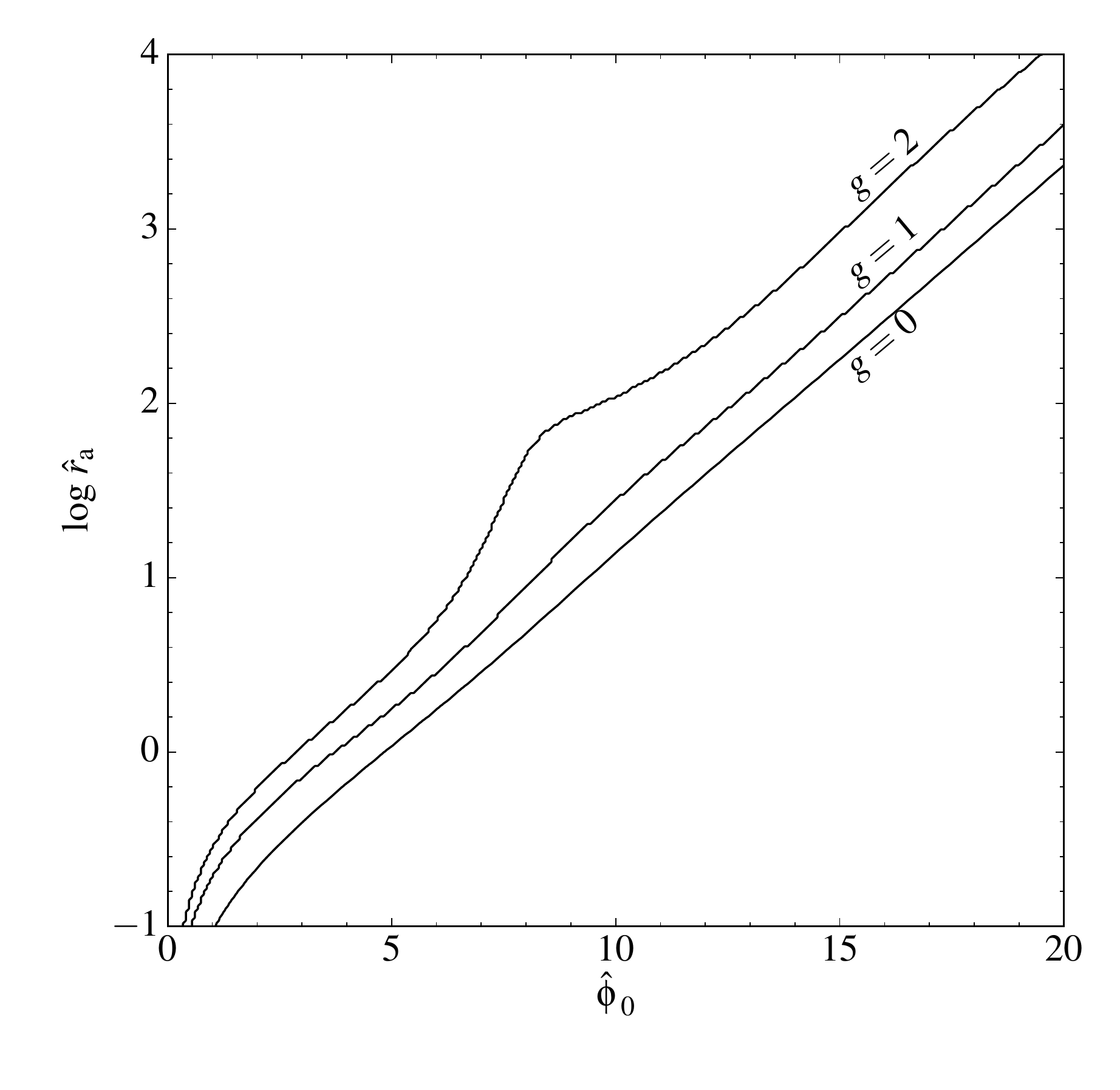}
 \caption{Minimum $\rahat$ for finite sized models, for different
 $\phihat_0$ and $g$.}
\label{fig:ramin}
\end{figure}

When considering anisotropic models, we find that for each $\phihat$
and $g$, there is a minimum value of $\rahat$ that can be used to
obtain a model that has a finite extent. We note that models with
infinite extent can have a finite total mass, but because we envision
an application of these models to tidally limited systems we do not
consider them here.  In Fig.~\ref{fig:ramin}, we show the minimum
$\rahat$ for which models are finite in extent.  The lines show, as a
function of $\phihat_0$, and for different $g$, the values of $\rahat$
that are needed to get $\rthat = 10^7$. Note that this minimum for
$\rahat$ goes up approximately exponentially with $\phihat_0$, and also increases
with $g$.

%______________________________________________________________________
\subsubsection{Entropy}
\citet{1966AJ.....71...64K} suggested that in the process of core
collapse, clusters evolve along a sequence of models with increasing
central concentration. He also noted that
his models are probably not able to describe the late stages of core
collapse, because for large central concentration the variation in
energy due to a change in the central concentration occurs in the
envelope, and not in the core. Further support for this idea comes
from \citet{1968MNRAS.138..495L}, who showed that a maximum in entropy
occurs at $\phihat_0 \simeq 9$ for both Woolley and King models at
constant mass and energy. The entropy of a self-gravitating system is
obtained from the DF as

\begin{equation}
S = -\int \dr^3 r \, \dr^3v \, f \ln f \ .
\label{eq:entropy}
\end{equation}
Because two-body encounters continuously increase the total entropy of
the system, we do not expect King models to be able to describe a
system in the late stages of core collapse
(i.e. $\phihat_0\gtrsim9$). This was confirmed by Fokker-Planck models
of isolated star clusters going into core
collapse \citep{1980ApJ...242..765C}, for which the entropy increase
follows that of King models with increasing central concentration, up
to a value of $\phihat_0 \simeq 9$, but then it continues to rise
during the gravothermal catastrophe. Cohn concluded that in this
regime, the isotropic King models are not able to describe the entropy
evolution in his simulations.

\begin{figure}
\includegraphics[width=\columnwidth]{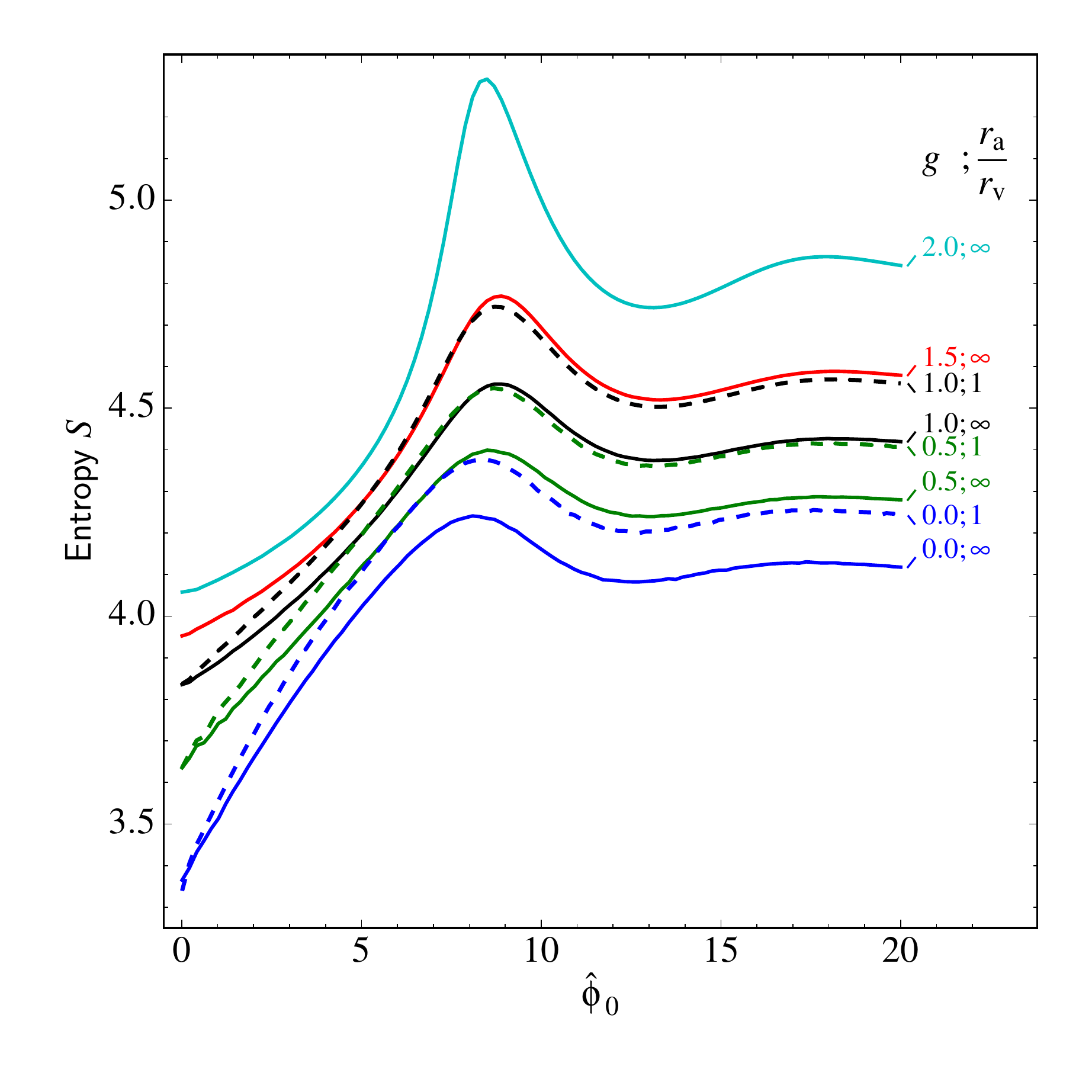}
 \caption{Entropy curves for isotropic and anisotropic models with
 different truncation prescriptions (i.e. different values of
 $g$). All models are scaled to the same mass and energy. The
 anisotropic models are shown as dashed lines and for these models we
 used $\ra=\rv$. For $g\ge1.5$ the anisotropic models are not finite
 for all $\phihat_0$, and the corresponding curves are therefore not
 plotted. This figure shows that the entropy can be increased by
 increasing $g$, and/or by decreasing $\ra$.}
\label{fig:entropy}
\end{figure}

In Fig.~\ref{fig:entropy}, we show the entropy $S$, computed as in
equation~(\ref{eq:entropy}), for the isotropic King models (black
solid line), which shows a maximum at $\phihat_0 \simeq 9$. We also
show the entropy curves for different values of $g$, and for selected
anisotropic models.  All models are scaled to the same $M$ and total
energy $\Etot$, in the conventional \henon\ $N$-body units:
$G=M=-4\Etot=1$ \citep{1971Ap&SS..14..151H}. For
$0\lesssim \phihat_0 \lesssim 1$, the anisotropic models are similar
to their corresponding isotropic models, and therefore they have
similar entropy. From this plot it is apparent that evolution at
constant mass and energy, and with increasing entropy is possible
beyond $\phihat\gtrsim9$ if $g$ is increased, and/or $\ra$ is
decreased (i.e. including more anisotropy).  A local maximum in
entropy is seen near $\phihat_0\simeq17$. Similar oscillating
behaviour of the entropy was found for isothermal models in a
non-conducting sphere and we refer to \citet{1968MNRAS.138..495L}
and \citet{1989ApJS...71..651P} for detailed discussions.  A study of 
equilibria in lowered isothermal models of the Woolley
and King-type can be found in \citet{1978ApJ...223..299K}; for a
discussion on the evolutionary sequence of quasi-equilibrium states in
$N$-body systems we refer to \citet{2005MNRAS.364..990T}. It would be
of interest to compare the models discussed here to the phase-space
density of particles in an $N$-body system undergoing core collapse.

\begin{figure}
\includegraphics[width=\columnwidth]{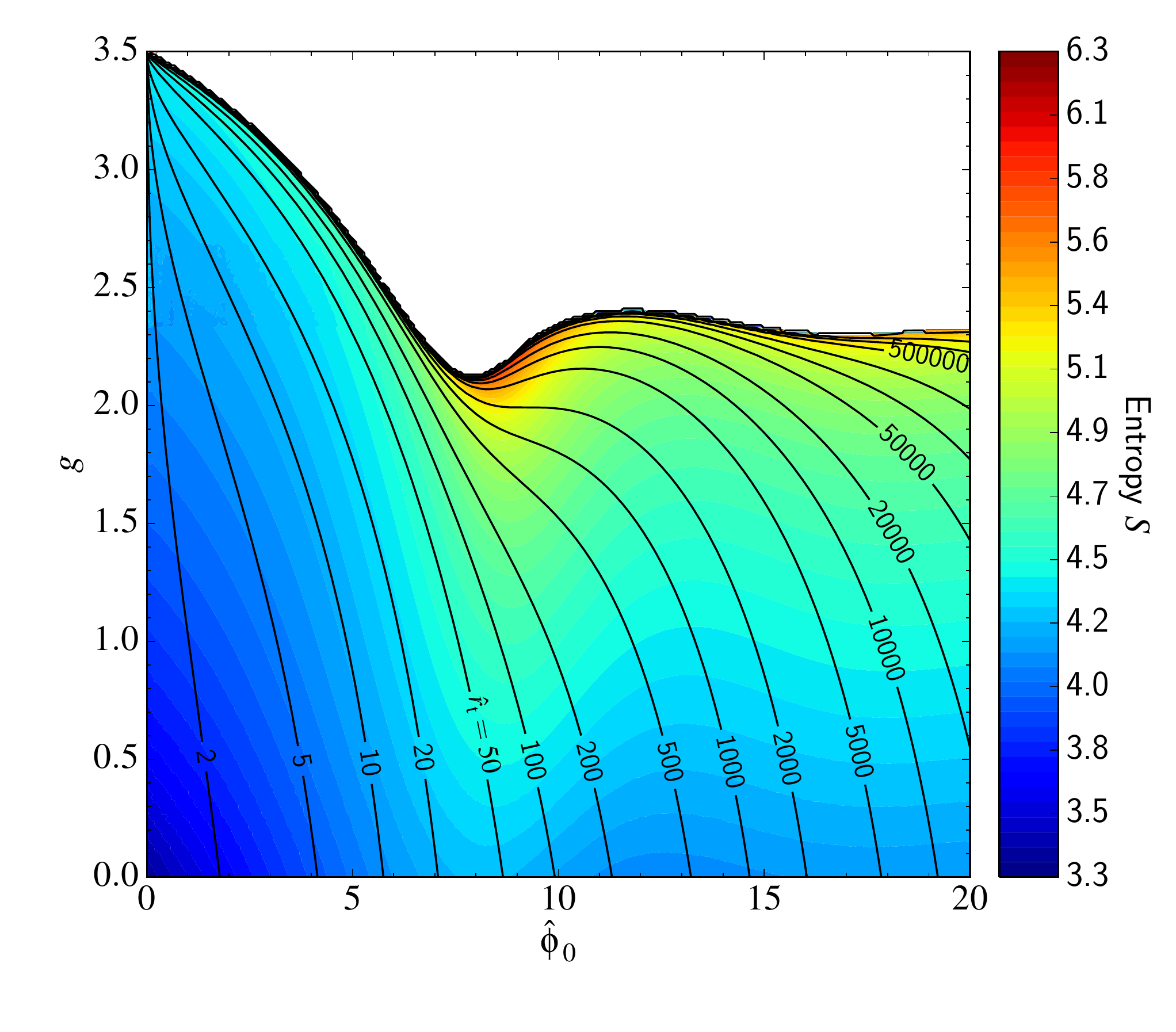}
 \caption{Entropy contours for isotropic models, all scaled to
 $G=M=-4\Etot=1$, with different $\phihat_0$ and $g$. Contours of
 constant $\rthat$ are shown as black lines. Moving to the right at
 constant $g$ leads to an increase of entropy up to
 $\phihat_0\simeq9$ \citep{1968MNRAS.138..495L,
 1980ApJ...242..765C}. The entropy can grow further by increasing $g$
 at constant $\phihat_0\simeq9$. The maximum entropy is found near
 $\phihat_0 \simeq9$ and $g\simeq2.2$.}
\label{fig:gphi0}
\end{figure}

In Fig.~\ref{fig:gphi0}, we illustrate the dependence of the entropy on
$g$ and $\phihat_0$ for isotropic models. For a model with $g=1$ and a
low concentration, the entropy can be increased by moving to the right
in this diagram, and near $\phihat_0\simeq9$ the entropy can be
increased by increasing $g$.

\begin{figure}
\includegraphics[width=\columnwidth]{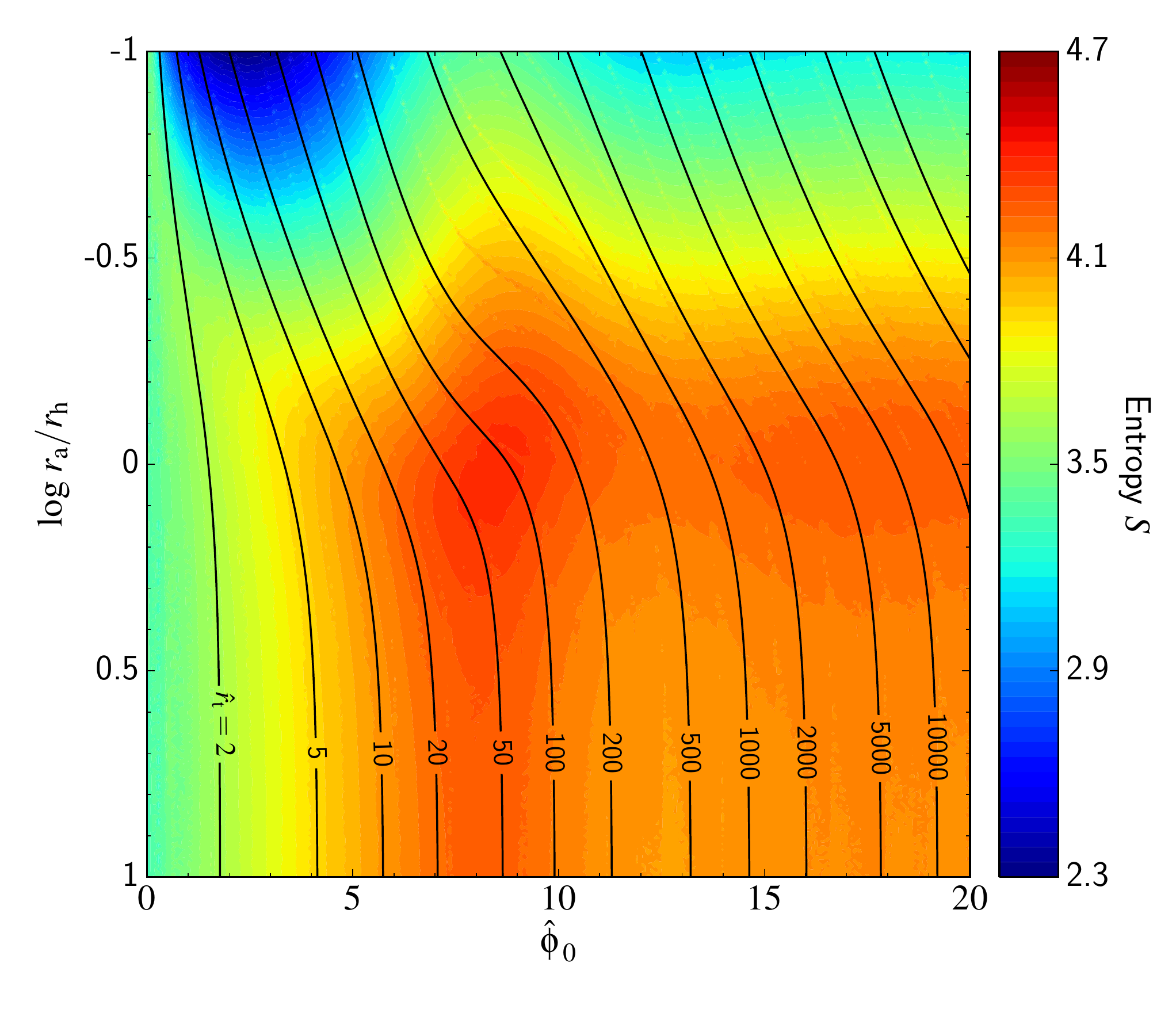}
 \caption{Entropy for models, all scaled to $G=M=-4\Etot=1$, with
 $g=0$ and different concentrations and different amounts of
 anisotropy, quantified here in terms of $\ra/\rh$. Contours of
 constant $\rthat$ are shown as black lines. A maximum in entropy is
 found at $\phihat_0\simeq9$ and $\ra\simeq\rh$.}
\label{fig:rarh_phi0}
\end{figure}

In Fig.~\ref{fig:rarh_phi0}, we show the dependence of entropy on
anisotropy, expressed here in terms of $\ra/\rh$, for models with
$g=0$. We see that for constant $\ra/\rh\gtrsim1$, the entropy can
increase by increasing $\phihat_0$, up to about
$\phihat_0\simeq9$ \citep[this was also found by][in a study of
anisotropic Woolley, King and Wilson models]{1998MNRAS.301...25M}. The
entropy can be increased further by decreasing the anisotropy
radius. A maximum is found near $\phihat_0\simeq9$ and $\ra\simeq\rh$.

%_______________________________________________________
\subsection{multimass models}
\label{ssec:multimass}
Multimass models with $\Ncomp$ mass bins require, in addition to the
parameters of the single-mass models, $2\Ncomp+2$ parameters
(Section~\ref{ssec:multimassmodel}).  There is, therefore, a large
variety of models that can be considered, and many properties that we
can chose from to illustrate the behaviour of these models.  We decide
to focus on two properties that highlight important features of these
multimass models in relation to mass segregation.  In a follow-up
study (Peuten et al., in preparation) we present a detailed comparison
between the multimass models and $N$-body simulations of clusters
with different mass functions.

\subsubsection{The role of $\delta$}
\label{ssec:delta}
\begin{figure}
\includegraphics[width=\columnwidth]{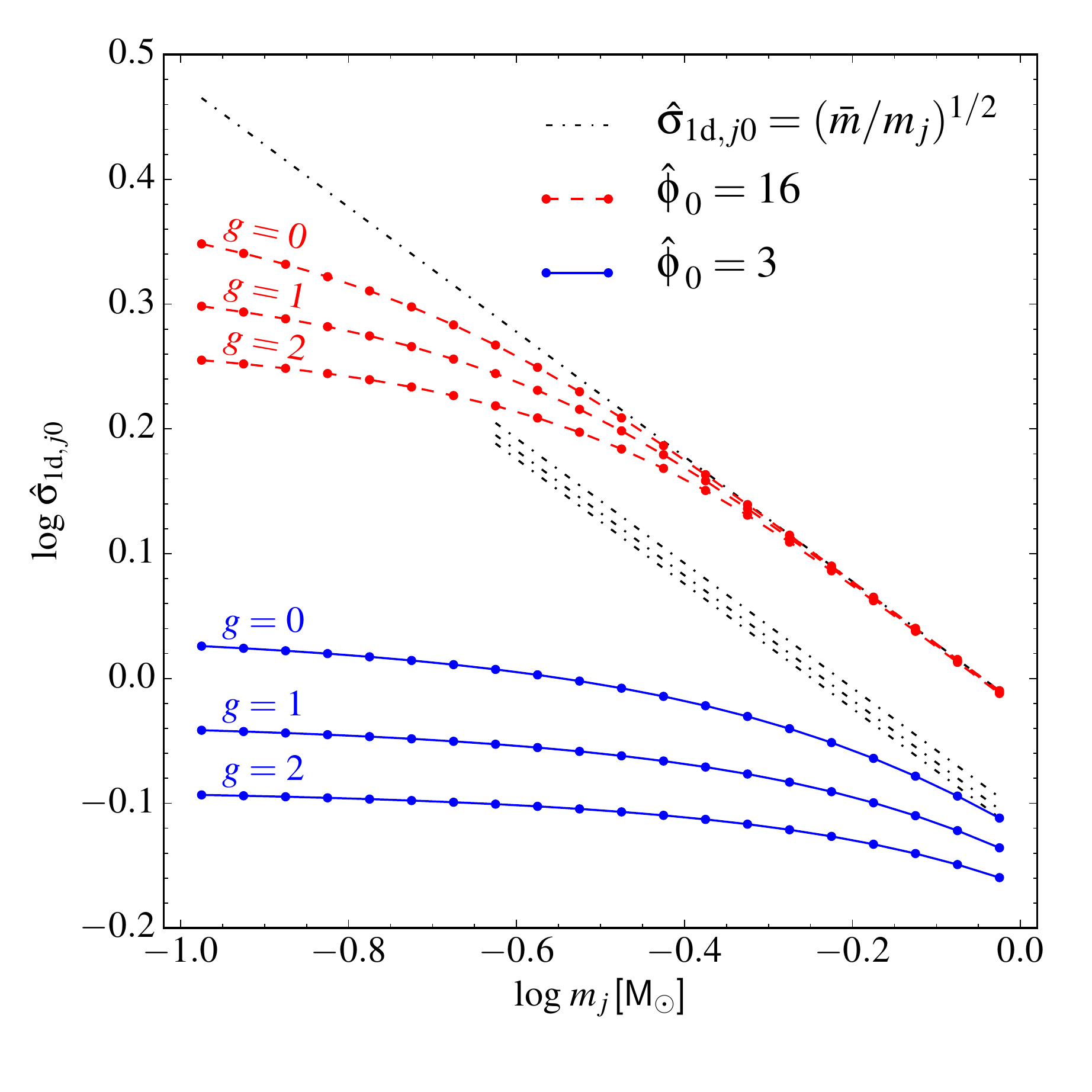}
 \caption{Dimensionless central velocity dispersion for each
 mass component of multimass models with a power-law mass function,
 and different $\phihat_0$ and $g$. The value of the mass segregation
 parameter is $\delta=1/2$. Equipartition in energy is only reached
 for large values of $m_j$ for the model with $\phihat_0=16$. The
 dash-dotted lines show the velocity dispersion each of the models
 would have in the case of equipartition.}
\label{fig:vrms0}
\end{figure}

In Fig.~\ref{fig:vrms0} we show the dimensionless central
velocity dispersion of each mass component, $\hat{\sigma}_{{\rm 1d},j0}$, as a
function of its mass $m_j$ for isotropic, 20-component models with
different $\phihat_0$ and $g$. The mass bins
are logarithmically spaced between $0.1\,\msun$ and $1\,\msun$  
(note that the units are not important, because the model behaviour depends only on the dimensionless values $m_j/\bar{m}$), and
$M_j\propto m_j^{0.7}$, which corresponds to a power-law mass function
$\dr N/\dr m_j \propto m_j^{-1.3}$ (i.e. a GC-like mass function). The
mass segregation parameter was set to $\delta=1/2$ (for the definition
of $\delta$, see equation~\ref{eq:delta}).

Despite the fact that $m_j s_j^2$ is constant for all mass bins, there
is no equipartition between the different mass species,
i.e. $\sigmaonedj$ does not scale as $m_j^{-1/2}$ for the different
mass components. This is because only in the limit of infinite central
concentration $\phihat_0\rightarrow\infty$, $s_j = \sigmaonedj$, but
for realistic values of $\phihat_0$, the ratio
$\sigmaonedj/s_j<1$. Because the central potential for the lower mass
components is smaller than the global $\phihat_0$ that defines the
model, the truncation in energies reduces $\sigmaonedj$ more for
low-mass components \citep{1981AJ.....86..318M,
2006MNRAS.366..227M}. This is illustrated by the $\phihat_0=16$ model
in Fig.~\ref{fig:vrms0}, for which a constant $m_j\sigmaonedj^2$ only
holds for the most massive bins.

\citet{2013MNRAS.435.3272T} recently observed very similar trends
between $\sigmaonedj$ and $m_j$ in $N$-body models of GCs (see their
fig.~1) as those shown in Fig.~\ref{fig:vrms0}. They concluded that
modelling techniques that assume equipartition, such as multimass
Michie-King models, are `approximate at best'. We stress that
multimass models that are widely used in literature, i.e. those with
$\delta=1/2$ \citep{1976ApJ...206..128D, 1979AJ.....84..752G}, are
{\it not} in a state of equipartition, as is illustrated in
Fig.~\ref{fig:vrms0} and has been stated
previously \citep{1981AJ.....86..318M, 2006MNRAS.366..227M}. In fact,
from a comparison of the model behaviour in Fig.~\ref{fig:vrms0} and
the $N$-body models of Trenti and van der Marel we conclude that the
most commonly chosen flavour of multimass models (i.e. King models
with $\delta = 1/2$) do a good job in reproducing the degree of mass
segregation in evolved stellar system \citep[see also][]{sollima15}.

%_____________________________
\subsubsection{The role of $\eta$}
\begin{figure}
\includegraphics[width=\columnwidth]{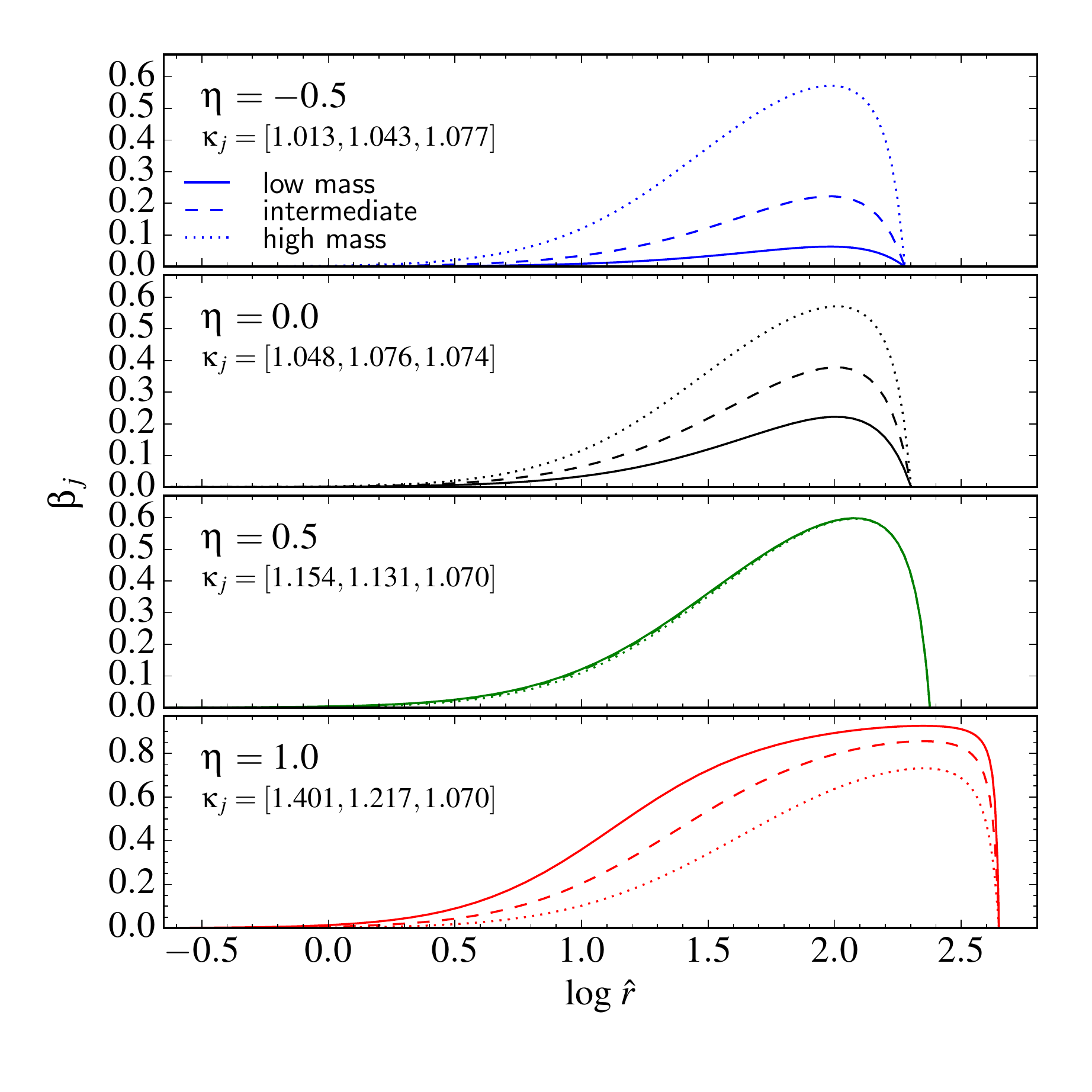}
 \caption{Anisotropy profiles for three-component models ($m_j = [0.2,
 0.4, 0.8]$) with different values for the anisotropy parameter
 $\eta$, that sets the anisotropy radius of the individual components
 as a function of their mass. All models have $\phihat_0=9$, $g=1.5$,
 $\delta=1/2$ and $\rahat=20$. The values of $\kappa_j$ are shown for
 each component in the individual panels.}
\label{fig:beta}
\end{figure}

In Fig.~\ref{fig:beta}, we illustrate the effect of the parameter
$\eta$ that sets the anisotropy radius of the different mass components (for
the definition of $\eta$ see equation~\ref{eq:eta}). We show the
anisotropy profiles for three-component models with $m_j = [0.2, 0.4,
0.8]$, and the same mass function as before (i.e. $M_j \propto
m_j^{0.7}$), and for different values of $\eta$. All models have
$\phihat_0=9$, $g=1.5$, $\delta=1/2$ and $\rahat = 20$.

In the multimass models used in the literature $\eta$ is implicitly
assumed to be $0$. From Fig.~\ref{fig:beta} we can see that this
implies that the $\beta$ profile of the high-mass stars rises to
larger values. It is tempting to interpret this as that massive stars
are on more radial orbits. However, the more massive stars are also
more centrally concentrated, where the velocity distribution is
more isotropic. To quantify the importance of this effect, we show in
each panel the values of the parameter $\kappa_j$ for each mass
component (equation~\ref{eq:kappa}). From this we can see that in fact
for the $\eta=0$ models the intermediate mass component is the most
anisotropic. The relation between $\beta_j$ and $\kappa_j$ depends on
the mass function, $\phihat_0$, and $g$ and this is therefore not a
general property of $\eta=0$ models.

We note that for $\eta=\delta=1/2$ the $\beta_j$ profiles are nearly
mass independent. Again, this does not mean that the kinetic
energy in radial orbits relative to that in tangential orbits is
constant, as can be seen from the values of $\kappa_j$. When
considering a value of $\eta > 1/2$ we observe that the component for
which $\beta_j$ assumes the largest values is the least massive one.

%%%%%%%%%%%%%%%%%%%%%%%%%%%%%%%%%%%%%%%%%%%%%%%%%%%%%%%%%%%%%%
\section{The \limepy\ code}
\label{sec:limepycode}

\subsection{General implementation}
\label{ssec:generalimplementation}
We introduce a \python-based code that solves the models and allows
the user to compute some useful quantities from the DF. The code is
called { Lowered Isothermal Model Explorer in PYthon} (\limepy),
and is available from: \url{https://github.com/mgieles/limepy}.

One of the main features of the code is its flexibility: the user can
easily solve isotropic or anisotropic models, and include one or more
mass components. The type of model to calculate is determined by the
input parameters:
\begin{enumerate}
\item the dimensionless central potential $\phihat_0$;
\item the truncation parameter $g$;
\item the anisotropy radius $\rahat$ (for anisotropic models);
\item two arrays  $m_j$, $M_j$ and $\delta$ and $\eta$ (for multimass models).
\end{enumerate}

By default, the model is solved in the dimensionless units described
in Section~\ref{ssec:scaling}.  There we pointed out that the scales
of the models are set by $A$ and $s$, which correspond to a mass
density (in six-dimensional phase space) and a velocity scale.  These two
scales, combined with the gravitational constant $G$ then define the
radial scale.  To allow a user to scale a model to physical
units, we decided to use the mass and radial scale as input,
and from this the velocity scale is computed internally. The
reason for this is that we foresee that an important application
of the code is to recover the GC mass and radius from a comparison of
the models to data. It is possible to scale
the model to physical units by specifying $M$ in $\msun$ and a radial
scale (either $\rv$ or $\rh$) in $\pc$. The resulting unit of velocity is then $\kms$, with
$G=0.004302\,\pc\,(\kms)^2/\msun$. Alternative units, such as
the \henon\ units $G=\rv=M=1$ \citep{1971Ap&SS..14..151H}, can be
considered by redefining the scales.  After solving the model, the
values of all typical radii are available: the King radius $\rs$, the
half-mass radius $\rh$, the truncation radius $\rt$, the anisotropy
radius $\ra$, and the virial radius $\rv$.

The code solves Poisson's equation with the `dopri5'
integrator \citep*{hairer1993solving}, which is a Runge-Kutta
integrator with adaptive step-size to calculate fourth and fifth order
accurate solutions. It is supplied by the \scipy\
sub-package \integrate. The relative and absolute accuracy parameters
are chosen as a compromise between speed and accuracy and can be
adjusted by the user. The basic version of the code allows us to obtain,
as a result of this integration, only the potential as a function of
radius. The full model solution contains, in addition to the
potential, the density, the radial and tangential components of the
velocity dispersion, the global velocity dispersion profile, and the
anisotropy profile (equation~\ref{eq:beta}). It is possible to use the
potential calculated in this way to compute the value of the DF as a
function of input $E$ (isotropic models), or $E$ and $J$ (anisotropic
models), or positions and velocities.

After solving a model, the code carries out a simple test to see
whether it is in virial equilibrium: $2\hat{K} - \Uhat = 0$, where
$\hat{K}$ is the dimensionless total kinetic energy (recall that
$\Uhat$ is defined to be positive, equation~\ref{eq:Uhat}). For models
that are infinite in extent, the solver stops at a large radius, the
virial equilibrium assertion fails and the lack of convergence is
flagged.

For multimass models the central densities of the components need to
be found by iteration
(Section~\ref{ssec:multimass}). \citet{1979AJ.....84..752G} proposed a
recipe in which the ratios of central densities over the total
density, $\alpha_j$, are set equal to $M_j/\sum_j M_j$ in the first
iteration. Because for $\delta>0$ the more massive components are more
centrally concentrated, the amount of mass in these components is
underestimated in the first iteration, while the mass in low-mass
stars is overestimated. After each iteration, $\alpha_j$ is multiplied
by the ratio of $M_j/M_j^\prime$, where $M_j^\prime$ is the array of
masses obtained in the previous step, and then normalized again. This
is repeated until convergence. However, we found that for models with
low $\phihat_0$ and a wide mass spectrum the mass function does not
always converge with this method. We found that multiplying $\alpha_j$
by $\sqrt{M_j/M_j^\prime}$, instead, is more reliable and results in a
similar number of iterations for models that do converge with the
method proposed by Gunn and Griffin.

Solutions are not numerically stable when considering large values of
the arguments of the hypergeometric functions. To stabilize the
calculations, we adopt the asymptotic behaviour of the hypergeometric
function $\hyp(1, b, -x)$ and $\hyp(2, b, -x)$ for $x\ge700$ (see
equations~\ref{eq:1f1asym1} and \ref{eq:1f1asym2}). For
multimass models with a wide mass spectrum (e.g. when stellar mass
black holes are considered in addition to the stellar mass function),
the central potential of the massive component can be too large for
the computation of $\rhoint(\mu^{2\delta}\phihat, \rhat)$ (see
equation~\ref{eq:rhointratio}) in the first iteration. We therefore
use the approximation $\rhohat_j
= \exp[\mu^{2\delta}(\phihat-\phihat_0)]$ if $\mu^{2\delta}\phihat
>700$.

%_________________________________________________________
\subsection{Model properties in projection}
\label{ssec:project}
In order to compare the models to observations of GCs, it is necessary
to compute the model properties in projection. For a spherically
symmetric system, it is straightforward to compute the projected
properties as a function of the projected radial coordinate
$R$ \citep[for a more detailed discussion, see for example][]{BT1987}.

The projected surface mass density is found from the intrinsic mass density as
\begin{equation}
\Sigma(R) = 2 \int_0^{\rt} \dr z \, \rho (r),
\label{Proj_Sigma}
\end{equation}
where $r^2 = R^2+z^2$, and $z$ is along the direction of the line-of-sight. 
The  velocity dispersion  along the line of sight is given by the following integral
\begin{equation}
\sigma_{\rm LOS}^2(R) = \frac{2}{\Sigma(R)} \int_0^{\rt} \dr z \, \rho (r) \sigma_z^2(r),
\label{Proj_Disp1}
\end{equation}
where $\sigma_z^2$ is the contribution of the velocity dispersion
tensor to the $z$-direction. For isotropic models, $\sigma_z^2
= \sigma^2/3$. For anisotropic models, it is possible to calculate
\begin{equation}
\sigma_z^2(r) = \sigma_{\rm r}^2 \cos ^2 \xi +  \sigma_{\theta}^2 \sin ^2 \xi,
\label{eq:sigma2zr}
\end{equation} 
where $\sin \xi = R/r$. We recall that, for the anisotropic models
considered here, $\sigma_{\theta}^2 = \sigma_{\varphi}^2
= \sigmat^2/2$. The component $\sigma_{\varphi}^2$ does not contribute
to $\sigma_z^2$, because it is always perpendicular to the line of
sight. We can use the anisotropy profile $\beta$ (see
equation~\ref{eq:beta}) to rewrite equation~(\ref{eq:sigma2zr})
as
\begin{equation}
\sigma_z^2(r) = \sigma_{\rm r}^2 \left[ 1 - \beta(r) \frac{R^2}{r^2} \right].
\end{equation} 

The quantity $\sigma_{\rm LOS}(R)$ is useful when comparing the models
to the velocity dispersion profiles that are calculated from radial
(i.e. line-of-sight) velocities. Now that proper motions data are
becoming available for an increasing number of
GCs \citep{2014ApJ...797..115B}, it is also interesting to compare the
velocity dispersion components that can be measured on the plane of
the sky with those calculated from the models. This comparison is
particularly important because it is a direct way to detect the
presence of anisotropy in the systems. We calculate, therefore, the
radial and tangential projected components of the velocity dispersion
as
\begin{align}
\sigma_{\rm R}^2(R) &= \frac{2}{\Sigma(R)}\int_0^{\rt} \dr z \, \rho (r) \sigma_{\rm S}^2(r), \\
\sigma_{\rm T}^2(R) &= \frac{2}{\Sigma(R)}\int_0^{\rt} \dr z \, \rho (r) \sigma_{\rm \varphi}^2(r),
\end{align}
where $\sigma_{\rm S}^2$ is given by
\begin{equation}
\sigma_{\rm S}^2(r) = \sigma_{\rm r}^2 \left[ 1 - \beta(r) \left(\frac{1 - R^2}{r^2}\right) \right].
\end{equation} 

In the case of multimass models, the projected quantities introduced
above are calculated separately for each mass component, by replacing
every quantity in the equations above with the respective $j$th
profile.

\subsection{Generating discrete samples from the DF}
A separate sampling routine {\textsc{limepy.sample}} is
provided that generates discrete samples from the models. The routine
takes a \python\ object containing a model as input and the number of
points $N$ that need to be sampled. In the case of a multimass models
the input $N$ is ignored and computed from the total mass $M$ and the
pair $m_j, M_j$. Radial positions are sampled by generating numbers
between 0 and 1 and interpolating the corresponding $r$ values from
the (normalized) cumulative mass profile(s).

To obtain velocities, we first sample values of $x$, where
$x=\khat^{3/2}=(\hat{v}^2/2)^{3/2}$. The probability density function
(PDF) for $x$ can be written as
\begin{equation}
P(x) = \frac{F(\phat x^{1/3})}{\phat
x^{1/3}}\Eg(g, \phihat(\rhat)-x^{2/3}),
\end{equation}
where $\phat=\rhat/\rahat$.  The function $P(x)$ has a maximum at
$x=0$, and declines monotonically to 0 at $x=\phihat(\rhat)^{3/2}$.
These properties make it easier to efficiently sample values for $x$
from $P(x)$, than sampling values of $v$ from $v^2 f(r,v)$. To make
the rejection sampling more efficient, we adapt a supremum function
$F(x)$, which consists of 10 segments between 11 values $x_i$ which
are linearly spaced between $0$ and $\phihat(\rhat)^{3/2}$, and for
each segment $x_i< x<x_{i+1}$, $F_i(x) = P(x_i)$. We then sample
values from the function $F(x)$, reject the points that are above
$P(x)$ and resample the rejected points until all points are
accepted. Typically, a handful of iterations are needed.

For anisotropic models we also need to sample angles $\theta$. We do
this by sampling values for $t=\cos\theta$. From the DF it follows
that the PDF for $t$ is
\begin{equation}
P(t) = \exp\left[\phat^2\khat(t^2-1)\right].
\label{eq:Pt}
\end{equation}
By integrating equation~(\ref{eq:Pt}) we find that the cumulative DF
is the imaginary error function. This function cannot be inverted
analytically, hence the values for $t$ need to be found by numerically
inverting this function, which can be done accurately with
built-in \scipy\ routines.

When values for $r, \vt$ and $\vr$ are obtained, these are converted
to Cartesian coordinates by generating three additional random angles.

%%%%%%%%%%%%%%%%%%%%%%%%%%%%%%%%%%%%%%%%%%%%
\section{Conclusions and discussion}
\label{sec:conclusion}

In this study we present a family of lowered isothermal models, with
the ability to consider multiple mass components and a variable amount
of radially biased pressure anisotropy. The models extend the
single-mass family of isotropic models recently developed by GV14. The
new additions we propose here make the models ideally suited to be
compared to data of resolved GCs.

The models are characterized by an isothermal and isotropic core, and
a polytropic halo. The shape of the halo is set by the truncation
parameter $g$, that controls the sharpness of the energy truncation,
i.e. the prescription of lowering the isothermal model. For integer
values of $g$, several well-known isotropic models are recovered: for
$g=0$ we recover the \citet{1954MNRAS.114..191W} models, for $g=1$
the \citet{1963MNRAS.125..127M}, or \citet{1966AJ.....71...64K} models
and $g=2$ corresponds to the non-rotating,
isotropic \citet[][]{1975AJ.....80..175W} models. The DF proposed by
GV14, with the introduction of the continuous parameter $g$ to
determine the truncation, allows us to consider models {\it in
between} these models. The advantage of this prescription for the
truncation is that it is now possible to control the sharpness of the
truncation by means of a parameter.

We present { Lowered Isothermal Model Explorer in PYthon}
($\limepy$), a \python-based code that solves the models, and computes
observable quantities such as the density and velocity dispersion
profile in projection. In addition, the code can be used to draw
random positions and velocities from the DF, which can be used to
generate initial conditions for numerical simulations.

It is interesting to discuss possible extensions of, and
improvements to the models. One obvious pitfall is that the tidal
field is not included in a self-consistent way. To quantify the effect
of the omission of the tidal field, we can consider the specific
energy $E$ at $\rt$. In our models $E(\rt)=\phi(\rt)=-GM/\rt$, whilst
inclusion of the tidal terms would give (for a cluster on a circular
orbit, in a reference frame corotating with the galactic orbit) a
specific Jacobi energy of $E_{\rm J}(\rt) = -(3/2)GM/\rt$. Therefore,
the properties of stars near the critical energy for escape are
described only approximately by these models, because in this energy
range the galactic tidal potential is of comparable importance as the
cluster potential. Another simplification of the models is that the
galactic tidal potential is triaxial, whilst our models are
spherical. Both of these points could be improved upon by including a
galactic tidal potential in the solution of Poisson's equation,
following the methods described by \citet{1995MNRAS.272..317H}
or \citet{2008ApJ...689.1005B}, \citet{2009ApJ...703.1911V}.

The models do not include a prescription for rotation, which can be an
important factor to take into account when describing real
GCs \citep[e.g.][]{2012A&A...538A..18B}. Self-consistent models with
realistic rotation curves exist \citep{2012A&A...540A..94V} and have
been successful in describing the rotational properties of several
Galactic GCs \citep{2013ApJ...772...67B}. It is feasible to include
rotation in the models presented in this paper, for example, in the
way it is done in the \citet[][]{1975AJ.....80..175W} model, by
multiplying the DF in equation~(\ref{eq:dfani}) by a $J_{\rm z}$
dependent exponential term. Including the rotation, and a description
of the galactic tidal field, would make the models more realistic and,
therefore, a worthwhile exercise for future studies.

Lastly, we note that our models could be useful in modelling
nuclear star clusters. Despite the fact that these systems are not
tidally truncated in the same way as clusters on an orbit around the
galaxy centre, their profiles are well described by lowered isothermal
models \citep[e.g.][]{2014MNRAS.441.3570G}. For a general application
to nuclear star clusters, it is desirable to include the effect of the
presence of a black hole in the centre, which generates a point-mass
potential. \citet{2010A&A...514A..52M} provided a method to
self-consistently solve King models with an external point-mass
potential: this recipe could be used to include the effect of a
massive black hole in the models presented here, to make them more
versatile in describing nuclear star clusters.

The aim of this project was to introduce models that can be used to
describe the phase-space density of stars in tidally limited, 
mass-segregated star clusters, in any stage of their life-cycle. At early
stage, GCs are dense with respect to their tidal
density \citep[e.g.][]{2013MNRAS.432L...1A} and at the present day
about half of the GCs is still much denser than their tidal
density \citep{2010MNRAS.401.1832B, 2011MNRAS.413.2509G}. These GCs
ought to have a population of stars with radial orbits in their
envelopes, either as a left-over of the violent relaxation process
during their formation \citep{1967MNRAS.136..101L}, and/or because of
two-body ejections from the core \citep{1972ApJ...173..529S}. In this
phase we expect models with high values of $g$, and small $\ra$ to
describe GCs well. These models can thus describe GCs with large
Jacobi radii, relative to $\rh$. This applies to a large fraction of
the Milky Way GC population, and these objects are beyond the reach of
King models \citep{2010MNRAS.401.1832B}. In later stages of evolution,
GCs will be more tidally limited, and isotropic, hence we expect $g$
to reduce and $\ra$ to increase during the evolution (up to a value
that practically corresponds to having isotropic models). Capturing
these variations in GCs properties with continuous parameters has the
advantage that these parameters can be inferred from data. This avoids
the need of a comparison of goodness-of-fit parameters of
different models.

When only surface brightness data are available, it is challenging to
distinguish between models with different truncation flavours and
pressure anisotropy, because their role has an impact mostly on the
low-density outer parts, far from the centre of the cluster, where
foreground stars and background stars are dominating. The addition of
kinematical data of stars in the outer region of GCs greatly aids in
discriminating between models, but this is challenging at
present. Precise proper motions ($\lesssim1\,\kms$) can be obtained
with the {\it Hubble Space Telescope}
\citep[{\it HST}; e.g.][]{2006ApJS..166..249M, 2015arXiv150200005W}, but
the field of view of {\it HST} limits observations to the central parts of
Milky Way GCs. Radial velocity measurements of stars in the outer
parts of GCs are expensive because of the contamination of non-member
stars \citep{2012ApJ...751....6D}. The upcoming data of the ESA-{\it Gaia}
mission will improve this situation: the availability of all-sky
proper motions and photometry measurements will facilitate membership
selection, and for several nearby GC the proper motions will be of
sufficient quality that they can be used for dynamical modelling and
to unveil the properties of the hidden low-energy
stars \citep{2012MNRAS.420.2562A,2013MmSAI..84...83P, sollima15}. The
models presented in this paper allow for higher level of inference of
physical properties of GCs from these upcoming data.

In two forthcoming studies we will compare the family of models to a
series of direct $N$-body simulations of the long term evolution of
single-mass star clusters \citep{2016MNRAS.462..696Z} and multimass
clusters (Peuten et al., in preparation) evolving in a tidal field.

\section*{Acknowledgements}
MG acknowledges financial support from the European Research Council
(ERC-StG-335936, CLUSTERS) and the Royal Society (University Research
Fellowship) and AZ acknowledges financial support from the Royal
Society (Newton International Fellowship). This project was initiated
during the {\it Gaia} Challenge
(\url{http://astrowiki.ph.surrey.ac.uk/dokuwiki}) meeting in 2013
(University of Surrey) and further developed in the follow-up meeting
in 2014 (MPIA in Heidelberg). The authors are grateful for interesting
discussions with the {\it Gaia} Challenge participants, in particular
Antonio Sollima, Anna Lisa Varri, Vincent H\'{e}nault-Brunet, and
Adriano Agnello. Miklos Peuten and Eduardo Balbinot are thanked for
doing some of the testing of \limepy\ and Maxime Delorme for
suggestions that helped to improve the code. We thank Giuseppe
Bertin for comments on an earlier version of the manuscript and the
anonymous referee for constructive feedback. Our model is written in
the \python\ programming language and the following open source
modules are used for the \limepy\ code and for the analyses done for
this
paper: \numpy\footnote{\url{http://www.numpy.org}}, \scipy\footnote{\url{http://www.scipy.org}}, \matplotlib\footnote{\url{http://matplotlib.sourceforge.net}}.

\appendix
%%%%%%%%%%%%%%%%%%%%%%%%%%%%%%%%%%%%%%%%%%%%%%%%%%%%%%%%%%%%%%%%%%%%%
\section{Derivations}
\label{app:series}

The DF introduced in equation~(\ref{eq:dfani}) can be expressed as a
function of the dimensionless quantities $\phihat$, $\khat$ and
$\phat$ defined in Sections~\ref{ssec:scaling} and \ref{ssec:ani} as
\begin{equation}
f = A  \exp\left(-\khat \phat^2 \sin^2\theta\right)\Eg\left(g, \phihat - \khat\right).
\end{equation}
We want to calculate, for these models, the density and velocity
dispersion components. We recall that these quantities can be obtained
from the DF in the following way:\footnote{As noticed already in
Section~\ref{ssec:iso}, equation~(\ref{DF_vel_disp_i}) holds because
the mean velocity for the systems described by these models is zero
everywhere.}
\begin{align}
\rho &= \int \dr^3 v \, f,  \\
\sigma^2_{i} &= \frac{1}{\rho} \int \dr^3 v \, f  v^2_i, \label{DF_vel_disp_i}
\end{align}
where the subscript $i$ denotes the $i$-th component of the velocity
vector.  To carry out these integrals of the DF in the
three-dimensional velocity volume we can use the dimensionless
variable $\khat$ and the variable $t = \cos \theta$
\begin{equation}
\dr^3 v = \dr v \dr\theta \dr\varphi \,  v^2 \sin\theta = - \dr \khat  \dr t \dr\varphi \, \sqrt{\khat} (2 s^2)^{3/2}.
\end{equation}
In calculating the relevant quantities mentioned above, we encounter
the following integrals with respect to the variable $t$
\begin{align}
&\int_0^{1} \dr t \, \exp\left[-\khat \phat^2 (1-t^2)\right]
  = \frac{F\left(\sqrt{\khat}\phat\right)}{\sqrt{\khat}\phat} , \label{Itheta1} \\
&\int_0^{1} \dr t \, t^2 \exp\left[-\khat \phat^2 (1-t^2)\right]
  = \frac{1}{2 \khat\phat^2} - \frac{F\left(\sqrt{\khat}\phat\right)}{2 (\sqrt{\khat}\phat)^3}, \label{Itheta2} 
\end{align}
where $F(x)$ is the Dawson integral, whose properties are presented in
Section~\ref{App:Dawson}.  We use the above results to proceed and
derive the density and velocity components.

%_____________________________________________
\subsection{Density profile}
The density is calculated as
\begin{align}
\rho &= \int \dr^3v \, f   \nonumber \\
&= \frac{\tilde{A}}{\sqrt{\pi}} \int_0^{\phihat} \dr\khat \, \int_0^{1} \dr t \, \khat^{1/2}\exp\left[\khat \phat^2 (t^2-1)\right] \Eg\left(g, \phihat-\khat\right)   \nonumber \\
&= \tilde{A} \frac{2}{\sqrt{\pi}}\int_0^{\phihat} \dr\khat \, \frac{F\left(\sqrt{\khat} \phat\right)}{\phat  }\Eg\left(g, \phihat-\khat\right) \nonumber\\
&= \tilde{A} \rhoint,
\end{align}
where we replaced $\Gamma(3/2)$ by $\sqrt{\pi}/2$, and we introduced
$\tilde{A} = A \left(2 \pi s^2 \right)^{3/2}$ and we solved the
integral over $t$ as shown in equation~(\ref{Itheta1}). The integral
$\rhoint$ can be solved by first doing an integration by parts (by
using the results in equations~\ref{Egintegral} and \ref{DawsDeriv})
and by then using the convolution formula of
equation~(\ref{EG_convolution}) and the recurrence relation of
equation~(\ref{RecurrencyEg}) in the following way:
\begin{align}
\rhoint &= \frac{2}{\sqrt{\pi}}\int_0^{\phihat} \dr\khat \, \Eg\left(g, \phihat-\khat\right) \frac{F\left(\sqrt{\khat} \phat\right)}{\phat} \label{Irho} \\
&= \frac{2}{\sqrt{\pi}}\int_0^{\phihat} \dr\khat \, \Eg\left(g+1, \phihat-\khat\right)\times\left[ \frac{1}{2\sqrt{\khat}}  - \phat F\left(\sqrt{\khat} \phat\right) \right]\nonumber\\
&= \Eg\left(g+\fthree, \phihat\right)
- \phat^2 \frac{2}{\sqrt{\pi}}\int_0^{\phihat} \dr\khat \, \Eg\left(g, \phihat-\khat \right) \frac{F\left(\sqrt{\khat} \phat\right)}{\phat}  \nonumber \\
& \hspace{1cm} +\phat^2 \frac{2}{\sqrt{\pi}}\int_0^{\phihat} \dr\khat \, \frac{\left(\phihat-\khat\right)^g}{\Gamma(g+1)} \frac{F\left(\sqrt{\khat} \phat\right)}{\phat} \nonumber \\
&= \Eg\left(g+\fthree, \phihat \right) - \phat^2 \rhoint + \phat^2 \fintone.
\end{align}
The integral $\fintone$ can be calculated by substituting the Dawson
function for its series representation (see
equation~\ref{DawsonSeries}), by changing variable to $y
= \khat/\phihat$, by using the Beta function of equation~(\ref{Beta}),
and by recognizing the expression in equation~(\ref{1F1DefSer})
\begin{align}
\fintone &= \frac{2}{\sqrt{\pi}}\int_0^{\phihat} \dr\khat \, \frac{\left(\phihat-\khat\right)^g}{\Gamma(g+1)} \frac{F\left(\sqrt{\khat} \phat\right)}{\phat } \nonumber \\
&=  \frac{\phihat^{g+\frac{3}{2}}\hyp\left(1,g+\ffive,-\phat^2\phihat\right)}{\Gamma\left(g+\frac{5}{2}\right)},
\label{IntegralDawson1}
\end{align}
where $ \hyp\left(a,b,x\right)$ is the confluent hypergeometric
function (see Section~\ref{1F1}). Therefore, we can finally write the
density integral as
\begin{equation}
\rhoint \!= \!\frac{\Eg(g+\fthree, \phihat )}{1+\rrahat^2} + \frac{\phat^2}{1+\phat^2}   \frac{\phihat^{g+\frac{3}{2}}\hyp(1,g+\frac{5}{2},-\phat^2\phihat)}{\Gamma\left(g+\frac{5}{2}\right)}.
\label{DENSITY_Anis}
\end{equation}

%__________________________________________
\subsection{Velocity dispersion profiles}
\label{sapp:sigma_deriv}
The velocity dispersion profile can be computed in a similar way as
the density, by using again the result in equation~(\ref{Itheta1})
\begin{align}
\sigma^2\rho &= \int \dr^3v \, v^2 f \nonumber \\ 
&= \frac{2\tilde{A} s^2}{\sqrt{\pi}} \int_0^{\phihat} \dr\khat\!\int_0^{1} \dr t \, \exp\left[{-\khat \phat^2 (1-t^2)}\right] \khat^{3/2} \Eg(g, \phihat-\khat )  \nonumber \\
&= \frac{4  \tilde{A} s^2}{\sqrt{\pi} } \int_0^{\phihat} \dr\khat \,\frac{F(\sqrt{\khat} \phat)}{\phat} \khat  \Eg(g, \phihat-\khat )  \nonumber \\
&= \tilde{A} s^2 \vsqint.
\end{align}
The integral $\vsqint$ can be solved with an integration by parts,
then by using equation~(\ref{EG_convolution}), and finally, after
having used the recurrence relation of equation~(\ref{RecurrencyEg}),
by recognizing the presence of the integral $\rhoint$ found when
calculating the density
\begin{align}
\vsqint &= \frac{4}{\sqrt{\pi}}\int_0^{\phihat} \dr\khat \, \khat  \Eg\left(g, \phihat-\khat \right) \frac{F\left(\sqrt{\khat} \phat\right)}{\phat  }\label{DefVSQint}\\
&= \frac{4}{\sqrt{\pi}}\int_0^{\phihat} \dr\khat \, \Eg\left(g+1, \phihat-\khat \right)   \nonumber \\
& \hspace{1cm}\times \left[\frac{\sqrt{\khat}}{2} + \frac{F\left(\sqrt{\khat} \phat\right)}{\phat } -\khat\phat^2 \frac{F\left(\sqrt{\khat} \phat\right)}{\phat} \right] \nonumber \\
&= \Eg\left(g+\ffive, \phihat \right) + 2  \rhoint_{g+1} + \phat^2 \left(\finttwo - \vsqint \right),
\end{align}
where $\rhoint_{g+1} = \rhoint(g+1, \rrahat, \phihat)$.
We then solve the integral $\finttwo$ in a similar way as we did for $\fintone$, to get
\begin{align}
\finttwo &=  \frac{4}{\sqrt{\pi}}\int_0^{\phihat} \dr\khat \, \khat   \frac{F\left(\sqrt{\khat} \phat\right)}{\phat } \frac{\left(\phihat-\khat\right)^g}{\Gamma(g+1)} \nonumber \\
&= \frac{\phihat^{g+\frac{5}{2}}}{\Gamma\left(g+\fseven\right)} \left[\hyp\left(1,g+\fseven,-\phat^2\phihat\right) \right. \nonumber \\
& \left. +  2 \hyp\left(2,g+\fseven,-\phat^2\phihat\right) \right].
\label{IntegralDawson2}
\end{align}
Therefore, we finally have
\begin{align}
\sigma^2\rho &= \frac{\tilde{A}  s^2}{(\phat^2+1)} \left\lbrace \Eg\left(g+\ffive, \phihat \right)\left(\frac{3+\phat^2}{1+\phat^2}\right)   \right. \nonumber \\
& \left. + \frac{\phat^2  \phihat^{g+\ffive}}{\Gamma\left(g+\frac{7}{2}\right)} \left[ 2 \,  \hyp\left(2,g+\fseven,-\phat^2\phihat\right) \right.\right.  \nonumber \\
& \left.\left. + \hyp\left(1,g+\fseven,-\phat^2\phihat\right)  \left(\frac{3+\phat^2}{1+\phat^2} \right)  \right] \right\rbrace.
\label{DISP_Anis}
\end{align}
The radial component of the velocity dispersion is given by
\begin{align}
\sigmar^2\rho &= \int \dr^3 v \, (v \cos\theta)^2 f \nonumber \\
&= \frac{2\tilde{A}  s^2}{\sqrt{\pi}} \int_0^{\phihat} \dr\khat \, \khat^{3/2} \Eg\left(g, \phihat-\khat \right) \left[ \frac{1}{\khat\phat^2} - \frac{F\left(\sqrt{\khat}\phat\right)}{(\sqrt{\khat}\phat)^3} \right] \nonumber \\
&= \frac{\tilde{A}  s^2}{\phat^2} \left[ \Eg\left(g+\fthree, \phihat \right) - \rhoint \right],
\label{DISP_r_Anis_int}
\end{align}
where we solved the integral by using equations~(\ref{Itheta2}) and
(\ref{EG_convolution}). We point out that we can express this quantity
by means of the density integrals of the isotropic and the anisotropic
case (see also equations~\ref{eq:rhointiso} and \ref{eq:rhointani} in
Section~\ref{sec:model}). We finally obtain
\begin{equation}
\sigmar^2\rho = \tilde{A}  s^2 \left[ \frac{\Eg\left(g+\fthree, \hat{\phi} \right)}{(1+\hat{p}^2)} - \frac{\hat{\phi}^{g+\fthree} \hyp\left(1,g+\ffive,-\hat{p}^2\hat{\phi}\right)}{(1+\hat{p}^2) \Gamma\left(g+\frac{5}{2}\right)} \right],
\label{DISP_r_Anis}
\end{equation}
which, by using equations~(\ref{RecurrencyEg}) and (\ref{RecurrencyHyp2}), can be rewritten as
\begin{equation}
\thickmuskip=2mu
\medmuskip=2mu
\sigmar^2\rho = \tilde{A}  s^2 \left[ \frac{\Eg\left(g+\ffive, \hat{\phi} \right)}{(1+\hat{p}^2)} + \frac{\phat^2 \hat{\phi}^{g+\ffive} \hyp\left(1,g+\frac{7}{2},-\hat{p}^2\hat{\phi}\right)}{(1+\hat{p}^2) \Gamma\left(g+\frac{7}{2}\right)} \right].
\label{DISP_r_Anis_REW}
\end{equation}
To calculate the tangential component of the velocity dispersion we
solve the integral over $t$ by expressing it as the difference between
equation~(\ref{Itheta1}) and equation~(\ref{Itheta2}), and we carry
out an integration by parts
\begin{align}
\sigmat^2\rho &= \int \dr^3 v \, (v \sin\theta)^2 f \nonumber \\
&= \frac{2 \tilde{A} s^2}{\sqrt{\pi}} \int_0^{\phihat} \dr\khat \, \khat^{3/2} \Eg\left(g, \phihat-\khat \right) \nonumber\\
&\hspace{0.25cm}\times\left[\frac{2 F\left(\sqrt{\khat}\phat\right)}{\sqrt{\khat}\phat} - \frac{1}{\khat\phat^2} +  \frac{F\left(\sqrt{\khat}\phat\right)}{(\sqrt{\khat}\phat)^3} \right] \nonumber \\
&= \frac{\tilde{A} s^2}{\phat^2} \left[\phat^2 \vsqint - \Eg\left(g+\fthree, \phihat \right) + \rhoint  \right] .
\label{DISP_T_Anis_int}
\end{align}
After recognizing the integrals we solved above, we can finally write
\begin{align}
\sigmat^2\rho &= \frac{\tilde{A} s^2}{(1+\phat^2)} \left\lbrace \Eg\left(\gfive, \phihat \right)\frac{2}{(1+\phat^2)}  \right.  \nonumber \\
& \left. + \, \frac{2\phat^2 \phihat^{g+\ffive}}{\Gamma(g+\fseven)} \left[  \frac{\hyp\left(1, g+\fseven, -\phihat\phat^2\right)}{(1+\phat^2)}  \right.\right.  \nonumber \\
& \left.\left. + \, \hyp\left(2, g+\fseven, -\phihat\phat^2\right) \right] \right\rbrace \ .
\label{DISP_T_Anis_REW}
\end{align}

%%%%%%%%%%%%%%%%%%%%%%%%%%%%%%%%%%%%%%%%%%%%%%%%%%%%%%%%%%%%%%%%%%%
\section{Derivations using fractional calculus}
\label{app:fractional}
Fractional calculus is a branch of mathematics that considers real
numbers for the orders of derivatives and integration. Because the
integrals that need to be solved contain terms like $\khat^{1/2}$ and
$\khat^{3/2}$, we can use semi-derivatives and semi-integrals and
integration by parts to solve them.  By
following \citet{2014arXiv1402.0319B}, we define the left- and
right-sided Riemann-Liouville fractional integrals of order $\alpha >
0$ of a function $q \in L^1$ as
\begin{align}
I^{\alpha}_{a+} q(t) &= \frac{1}{\Gamma(\alpha)} \int_{a}^{t} \dr x \, q(x) (t-x)^{\alpha-1}, \label{leftFracInt} \\
I^{\alpha}_{b-} q(t) &= \frac{1}{\Gamma(\alpha)} \int_{t}^{b} \dr x \, q(x) (x-t)^{\alpha-1}.  \label{rightFracInt}
\end{align}
In the remainder of this section, we will use the result illustrated
by \citet{2014arXiv1402.0319B} in their Proposition 2 \citep[for a
proof see ][]{samko1993}
\begin{align}
\int_{a}^{b} \dr t \, \left(I^{\alpha}_{a+} q_1\right)(t)  q_2(t) = \int_{a}^{b} \dr t \, q_1(t)  \left(I^{\alpha}_{b-} q_2\right)(t).
\label{Prop2}
\end{align}

%___________________________________________________
\subsection{Density}
When considering the isotropic limit of the DF, the integral to be
solved to calculate the density is:
\begin{equation}
\rhoint = \frac{2}{\sqrt{\pi}} \int_0^{\phihat}\dr\khat\, \khat^{1/2} \Eg\left(g, \phihat-\khat \right).
\label{densISOFrac1}
\end{equation}
We can use fractional calculus to solve it, by changing variable of
integration (using $x = \phihat-\khat$) and by considering that
\begin{align}
q_1(x) &= \Eg\left(g, x \right), \\
q_2(x) &= 1, \\
I^{1/2}_{0+} q_1(x) &= \Eg\left(g + \fone, x \right), \\
I^{1/2}_{\phihat-} q_2(x) &= \frac{2(\phihat-x)^{1/2}}{\sqrt{\pi}},
\end{align}
thus obtaining
\begin{align}
\rhoint &= \int_0^{\phihat}\dr x\, \Eg(g +\fone, x) = \Eg(g +\fthree, \phihat),
\end{align}
which was solved by using equation~(\ref{Egintegral}).

The density of the anisotropic models is calculated by solving the
integral $\rhoint$ introduced in equation~(\ref{Irho}). To do this, we
can use the result shown in equation~(\ref{Prop2}) by noticing (see
equations \ref{EG_convolution} and \ref{DawsonDef}) that
\begin{align}
q_1(\khat) &= \exp\left(-\khat\phat^2\right), \label{q11}\\
q_2(\khat) &= \Eg\left(g, \phihat-\khat\right), \\
I^{1/2}_{0+} q_1(\khat) &= \frac{2F(\sqrt{\khat}\phat)}{\sqrt{\pi} \phat}, \label{q1i1} \\
I^{1/2}_{\phihat-} q_2(\khat) &= \Eg\left(g + \fone, \phihat-\khat \right),
\end{align}
and by rewriting the integral of equation~(\ref{Irho}) as
\begin{equation}
\rhoint = \int_0^{\phihat} \dr \khat \, \exp\left(-\khat\phat^2\right) \Eg\left(g + \fone, \phihat-\khat \right).
\label{densFrac1}
\end{equation}
This integral can be solved by parts, by using the expression of the
derivative of the lower incomplete gamma function
(equation~\ref{GammaDeriv}) and by using equation~(\ref{IntDefHyp}) to
obtain
\begin{equation}
\rhoint = \frac{\Eg\left(g+\fone, \phihat \right)}{1+\phat^2} -  \frac{\phihat^{g+\fone}\hyp(1,g+\fthree, -\phihat\phat^2)}{(1+\phat^2)\Gamma(g+\fthree)}.
\end{equation}
The last step can be rewritten as equation~(\ref{DENSITY_Anis}) by
using the recurrence relations shown in equations~(\ref{RecurrencyEg})
and (\ref{RecurrencyHyp2}).

\subsection{Velocity dispersion}
\label{sapp:sigma_deriv_frac}
In the isotropic limit of the DF, the velocity dispersion is
calculated by means of an integral with the same structure as the one
found in equation~(\ref{densISOFrac1}).  The velocity dispersion of
the anisotropic models is calculated by solving the integral $\vsqint$
introduced in equation~(\ref{DefVSQint}). We can use the result shown
in equation~(\ref{Prop2}) also in this case, by considering the
function $q_1$ introduced in equation~(\ref{q11}), its fractional
integral (equation~\ref{q1i1}), and
\begin{align}
\thickmuskip=2mu
\medmuskip=2mu
q_2(\khat) &= \khat \, \Eg\left(g, \phihat-\khat \right), \\
I^{1/2}_{\phihat-} q_2(\khat) &= \frac{1}{2} \Eg\left(g+\fthree, \phihat-\khat \right) + \khat \Eg\left(g+\fone, \phihat-\khat \right),
\end{align}
and by rewriting the integral $\vsqint$ as
\begin{align}
\vsqint &= \int_0^{\phihat} \dr \khat \exp\left(-\khat\phat^2\right) \Eg\left(g +\fthree, \phihat-\khat \right)   \nonumber\\
& +2 \int_0^{\phihat} \dr \khat \exp\left(-\khat\phat^2\right) \khat \Eg\left(g +\fone, \phihat-\khat \right).
\end{align}
The first integral is in the same form as the one we found for the
density, and the second one can be reduced to something similar with
an integration by parts. By solving the integrals, we obtain:
\begin{align}
\vsqint &= \frac{1}{1+\phat^2 } \left[ \Eg\left(g+\fthree, \phihat\right)\left(\frac{3+\phat^2}{1+\phat^2}\right) - \frac{2 \phihat^{g+\fthree}}{\Gamma\left(g+\fthree\right)} \right.   \nonumber\\
& \left. +\frac{2 \phihat^{g+\fthree}\hyp(1,g+\frac{5}{2}, -\phihat\phat^2)}{\Gamma\left(g+\frac{5}{2}\right)} \left( g + \phat^2\phihat + \frac{\phat^2}{1+\phat^2} \right) \right],
\end{align}
which can be rewritten as equation~(\ref{DISP_Anis}) by using the
recurrence relations shown in equations~(\ref{RecurrencyEg}),
(\ref{RecurrencyHyp1}), and (\ref{RecurrencyHyp2}).

By inspecting equations~(\ref{DISP_r_Anis_int}) and
(\ref{DISP_T_Anis_int}), it is immediate to notice that the radial and
tangential components of the velocity dispersion are calculated with
integrals that can be written as a combination of those solved in this
section by means of fractional calculus.

%%%%%%%%%%%%%%%%%%%%%%%%%%%%%%%%%%%%%%%%%%%%%%%%%%%%%%%%%%%%%%%
\section{Differential energy distribution}
\label{app:dmde}

The differential energy distribution gives the amount of mass per
units of energy \citep{BT1987}. Here we briefly recall how to
calculate it for isotropic and anisotropic systems.

For a DF that only depends on $E$, the differential energy
distribution can be calculated as:
\begin{equation}
\frac{\dr M}{\dr E} \equiv f(E)\gdos(E),
\label{eq:dmde}
\end{equation}
where $f(E)$ is the DF, and $\gdos(E)$ is the density of states, which
is the volume of phase space per unit energy and is defined as
\begin{equation}
{\gdos}(E) = \int \dr^3r \, \dr^3v \, \delta(E-H),
\label{densityofstates}
\end{equation}
where $\delta(x)$ is the Dirac delta function. For a spherically
symmetric systems, this integral can be expressed as
\begin{equation}
\thickmuskip=2mu
\medmuskip=2mu
\gdos(E) = 16 \pi^2 \, \int_0^{\rmm(E)} \dr r \, r^2 \, \int \dr v \, v^2 \, \delta\left(\frac{1}{2} v^2 + \phi - E\right),
\label{densityofstates_1}
\end{equation}
where $\rmm(E)$ is the radius at which $\phi = E$. By changing
variable (using $y = v^2/2$), we finally obtain
\begin{equation}
\gdos(E) = 16 \pi^2 \, \int_0^{\rmm(E)} \dr r \, r^2 \, \sqrt{2(E - \phi)}, 
\label{densityofstates_2}
\end{equation}
and the differential energy distribution is therefore
\begin{equation}
\frac{\dr M}{\dr E} = 16 \pi^2 \, f(E) \, \int_0^{\rmm(E)} \dr r\, r^2 \, \sqrt{2(E-\phi)}.
\label{N_E_iso}
\end{equation}

When considering anisotropic systems, for which the DF depends also on
the angular momentum $J$, the differential energy distribution is
obtained as
\begin{equation}
\frac{\dr M}{\dr E} = \int \dr^3r \, \dr^3v \, \delta(E-H) f(H,J).
\label{differentialenergydistribution}
\end{equation}
This integral can be expressed as
\begin{equation}
\frac{\dr M}{\dr E}= 8 \pi^2 \int \dr r \, r^2 \int \dr \vr \, \dr \vt \, \vt \, \delta(E-H) f(H,J), 
\end{equation}
and it can then be rearranged by changing variable and introducing $J = r \, \vt$ in the following way
\begin{equation}
\frac{\dr M}{\dr E} = 8 \pi^2 \int \dr r \int \dr \vr \, \dr J \, J \, \delta(E-H) f(H,J).
\end{equation}
The integral over $\vr$ is solved by using the fact that
\begin{equation}
\vr^2 = 2(E-\Phi) - \frac{J^2}{r^2},
\end{equation} 
to obtain 
\begin{align}
\frac{\dr M}{\dr E} = 16 \pi^2 \, \int \dr r \int \dr J \, \frac{J f(E,J)}{\sqrt{2(E-\Phi) - J^2/r^2}}.
\end{align}
By using the expression for the DF of the models presented in this
paper (equation~\ref{eq:dfani}) we can perform the integration over
$J$ in this last equation and write the differential energy
distribution as a function of the part of the DF that depends on
energy only, $f(E)$:
\begin{align}
\frac{\dr M}{\dr E}= 16 \pi^2 \, f(E) \, \int_0^{\rmm(E)}\dr r\, \sqrt{2} \, \ra s \, r \, F\left(\frac{r \sqrt{E-\phi}}{\ra s}\right),
\end{align}
where $F(x)$ is the Dawson integral (see
Appendix~\ref{App:Dawson}). In the limit of $\ra\rightarrow\infty$
this reduces to the result for the isotropic case shown in
equation~(\ref{N_E_iso}), which follows from substituting the leading term of equation (D14) in equation (C11).

%%%%%%%%%%%%%%%%%%%%%%%%%%%%%%%%%%%%%%%%%%%%%%%%%%%%%%%%%%%%%%
\section{Useful properties of mathematical functions}
\label{app:functions}
\subsection{Useful properties of the gamma functions}
\label{app:gamma}
The gamma  function of a positive integer $n$ is defined as
\begin{equation}
\Gamma(n) = (n-1)!,
\end{equation}
while for non-integer arguments $a$, it can be written as an integral
\begin{equation}
\Gamma(a) = \int_0^{\infty} \dr t \, t^{a-1}\exp(-t).
\end{equation}
The lower incomplete gamma function is given by 
\begin{equation}
\gamma(a, x) = \int_0^x \dr t \, t^{a-1}\exp(-t),
\end{equation}
and its derivative is
\begin{equation}
\frac{\dr \gamma(a,x)}{\dr x} = x^{a-1}\exp(-x).
\label{GammaDeriv}
\end{equation}

\subsection{Useful properties of the $\Eg(a,x)$ function}
\label{AppD:Eg}
The exponential function $\Eg(a,x)$ is defined as
\begin{equation}
\Eg(a,x) = \frac{1}{\Gamma(a)} \int_0^x  \dr t \, t^{a-1} \exp(x-t),
\end{equation}
and an alternative expression is given by means of the lower
incomplete gamma function
\begin{equation}
\Eg(a,x) = \frac{\exp(x) \gamma(a,x)}{\Gamma(a)}. 
\label{E_gamma_exp}
\end{equation}
The series representation of this function is
\begin{equation}
\Eg(a,x) = \sum_{i=0}^{\infty} \frac{x^{i+a}}{\Gamma(i+a+1)}.
\label{Eg_series_repr}
\end{equation}
The following recurrence relation holds
\begin{equation}
\Eg(a,x) = \Eg(a+1, x) + \frac{x^a}{\Gamma(a+1)}.
\label{RecurrencyEg}
\end{equation}
The derivative and the integral of $\Eg(a,x)$ are given by
\begin{align}
\frac{\dr \Eg(a, x)}{\dr x} &= \Eg(a-1,x), \\
\int \dr x \, \Eg(a, x) &= \Eg(a+1,x) + \mathrm{constant}. \label{Egintegral}
\end{align}
A proof of these equations can be easily obtained by writing
$\Eg(a,x)$ as in equation~(\ref{E_gamma_exp}), and by considering
equation~(\ref{GammaDeriv}) and the recurrence relation
(equation~\ref{RecurrencyEg}).  The convolution formula
\begin{equation}
\frac{1}{\Gamma(b)} \int_0^x \dr y \, \Eg(a,x-y) y^{b-1} = \Eg(a+b,x)
\label{EG_convolution}
\end{equation}
can be obtained by using the series representation of $\Eg(a,x)$
(equation~\ref{Eg_series_repr}) and by changing variable, to express
the integral with a form that allows us to recognize the Beta function:
\begin{equation}
B(m,n) = \int_0^1 \dr y \, (1-y)^{m-1}y^{n-1} = \frac{\Gamma(m)\Gamma(n)}{\Gamma(m+n)}.
\label{Beta}
\end{equation}
The identity of equation~(\ref{EG_convolution}) accounts for the
fractional integration of $\Eg(a,x)$ (see equation~\ref{leftFracInt}).

\subsection{Useful properties of the Dawson integral}
\label{App:Dawson}
The Dawson integral (sometimes called the Dawson function) is defined as
\begin{equation}
F(x) = \exp(-x^2) \int_0^x \dr y \exp(y^2).
\label{DawsonDef}
\end{equation}
It is also possible to express $F(x)$ as a sum as
\begin{equation}
F(x) = \sum_{i=0}^{\infty} \frac{(-1)^i x^{2i+1} \Gamma\left(\fthree\right)}{\Gamma\left(i+\fthree \right)} \ .
\label{DawsonSeries}
\end{equation}
The Dawson integral is an odd function, and its derivative is 
\begin{equation}
\frac{\dr F(x)}{\dr x} = 1 - 2 x F(x).
\label{DawsDeriv}
\end{equation}

%__________________________________________________________________________
\subsection{Useful properties of the confluent hypergeometric function}
\label{1F1}
The confluent hypergeometric function is defined as
\begin{eqnarray}
\hyp(a,b,x) &= \displaystyle\sum_{i=0}^{\infty} \frac{\Gamma(a+i)}{\Gamma(a)}\frac{\Gamma(b)}{\Gamma(b+i)}\frac{x^i}{\Gamma(i+1)}.
\label{1F1DefSer}
\end{eqnarray}
It can also be defined by means of an integral, as
\begin{equation}
\hyp(a,b,x) = \frac{\Gamma(b)}{\Gamma(a) \Gamma(b-a)} \int_{0}^{1} \dr y \, \exp(xy) y^{a-1} (1-y)^{b-a-1},
\label{IntDefHyp}
\end{equation}
which holds for ${\rm Re}(b)>{\rm Re}(a)>0$. The following recurrence relations hold
\begin{align}
\hyp(2,b,x) &= (2-b+x) \, \hyp(1,b,x) + b - 1, \label{RecurrencyHyp1} \\
x \, \hyp(1,b+1,x) &= b \, \hyp(1,b,x) - b.  \label{RecurrencyHyp2}
\end{align}
We also note that this function is related to the exponential
function, and for $b=a$ we have $\hyp(a,a,x)=\exp(x)$. Another useful
property of this function is that
\begin{equation}
\hyp(a, b, 0) = 1.
\label{Hyp1}
\end{equation}

The density (equation~\ref{eq:rhointanires}) and the velocity
moments (equations~\ref{eq:vrsqintres} - \ref{eq:vsqintres}) of
anisotropic models are expressed by means of the function
$\hyp(a,b,x)$ with $a=1$. When considering $a = 1$ in
equation~(\ref{IntDefHyp}), we obtain 
\begin{align}
\hyp(1,b,x) &= \frac{(b-1)}{x^{b-1}} \, \exp(x) \, \gamma(b-1,x) .
\label{1F1_alt}
\end{align}
We point out that when $x<0$ two of the quantities appearing in
equation~(\ref{1F1_alt}) are imaginary. This is the reason why in
general this expression cannot be used to speed up the code by
expressing the equations mentioned above in a more compact way. When
considering integer values of $b$, however, we can simplify the
hypergeometric function, and express it by means of exponentials and
polynomials. The smallest value of $b$ we use is $g +\ffive$,
therefore $b$ assumes the smallest integer value when $g = \fone$ for
$b=3$ we calculate
\begin{align}
\hyp(1,3,x) &= \frac{2}{x^{2}} \left[ \exp(x) -1 -x\right].
\end{align}
Combined with the recurrence relation mentioned earlier, the results
of the anisotropic models and half-integer values of $g$ can be
expressed by means of these elementary functions.

The asymptotic series expansion for the confluent hypergeometric function when $|x| \rightarrow \infty$ is given by
\begin{align}
\hyp(a,b,x) & \propto \frac{\Gamma(b)}{\Gamma(b-a)} (-x)^{-a} \left[ 1 + \mathcal{O}\left( \frac{1}{z} \right) \right]  \nonumber \\
            & +\frac{\Gamma(b)}{\Gamma(a)} \exp(x) x^{a-b}  \left[ 1 + \mathcal{O}\left( \frac{1}{z} \right) \right].
\end{align}
This is useful to compute the density and velocity moments, because
for very large $|x|$ the evaluation of this function becomes
inaccurate in \python\ (see section~\ref{ssec:generalimplementation}).
In particular, the functions that are needed to compute these
quantities, $\hyp(1,b,-x)$ and $\hyp(2,b,-x)$, have the following
behaviour for large $|x|$:
\begin{align}
\lim_{x\rightarrow \infty}\hyp(1,b,-x) &= \frac{b-1}{x},\label{eq:1f1asym1}\\
\lim_{x\rightarrow \infty}\hyp(2,b,-x) &= \frac{(b-1)(b-2)}{x^2}.\label{eq:1f1asym2}
\end{align}

\end{document}